\documentclass[twocolumn,english,superscriptaddress,10pt,aps,prl]{revtex4-1}
\usepackage[T1]{fontenc}
\usepackage[latin9]{inputenc}
\usepackage{color}
\usepackage{babel}
\usepackage{amsmath}
\usepackage{graphicx}
\usepackage[unicode=true,pdfusetitle,
 bookmarks=true,bookmarksnumbered=false,bookmarksopen=false,
 breaklinks=true,pdfborder={0 0 1},backref=false,colorlinks=true]
 {hyperref}

\makeatletter
\usepackage{pdfpages}
\usepackage{pgffor}
\usepackage[caption=false]{subfig}
\AtBeginDocument{\let\LS@rot\@undefined}

\@ifundefined{showcaptionsetup}{}{%
 \PassOptionsToPackage{caption=false}{subfig}}
\usepackage{subfig}
\makeatother

\begin{document}

\title{Guiding and confining of light in a two-dimensional synthetic space
using electric fields}

\author{Hamidreza Chalabi}
\email{hchalabi@umd.edu}

\affiliation{Department of Electrical and Computer Engineering and Institute for
Research in Electronics and Applied Physics, University of Maryland,
College Park, Maryland 20742, USA}

\affiliation{Joint Quantum Institute, University of Maryland, College Park, Maryland
20742, USA}

\author{Sabyasachi Barik}
\email{sbarik@umd.edu}

\affiliation{Department of Electrical and Computer Engineering and Institute for
Research in Electronics and Applied Physics, University of Maryland,
College Park, Maryland 20742, USA}

\affiliation{Joint Quantum Institute, University of Maryland, College Park, Maryland
20742, USA}

\affiliation{Department of Physics, University of Maryland, College Park, Maryland
20742, USA}

\author{Sunil Mittal}
\email{mittals@umd.edu}

\affiliation{Department of Electrical and Computer Engineering and Institute for
Research in Electronics and Applied Physics, University of Maryland,
College Park, Maryland 20742, USA}

\affiliation{Joint Quantum Institute, University of Maryland, College Park, Maryland
20742, USA}

\author{Thomas E. Murphy}
\email{tem@umd.edu }

\affiliation{Department of Electrical and Computer Engineering and Institute for
Research in Electronics and Applied Physics, University of Maryland,
College Park, Maryland 20742, USA}

\affiliation{Department of Physics, University of Maryland, College Park, Maryland
20742, USA}

\author{Mohammad Hafezi}
\email{hafezi@umd.edu}

\affiliation{Department of Electrical and Computer Engineering and Institute for
Research in Electronics and Applied Physics, University of Maryland,
College Park, Maryland 20742, USA}

\affiliation{Joint Quantum Institute, University of Maryland, College Park, Maryland
20742, USA}

\affiliation{Department of Physics, University of Maryland, College Park, Maryland
20742, USA}

\author{Edo Waks}
\email{edowaks@umd.edu}

\affiliation{Department of Electrical and Computer Engineering and Institute for
Research in Electronics and Applied Physics, University of Maryland,
College Park, Maryland 20742, USA}

\affiliation{Joint Quantum Institute, University of Maryland, College Park, Maryland
20742, USA}

\affiliation{Department of Physics, University of Maryland, College Park, Maryland
20742, USA}
\begin{abstract}
Synthetic dimensions provide a promising platform for photonic quantum
simulations. Manipulating the flow of photons in these dimensions
requires an electric field. However, photons do not have charge and
do not directly interact with electric fields. Therefore, alternative
approaches are needed to realize electric fields in photonics. One
approach is to use engineered gauge fields that can mimic the effect
of electric fields and produce the same dynamical behavior. Here,
we demonstrate such an electric field for photons propagating in a
two-dimensional synthetic space. We achieve this using a linearly
time-varying gauge field generated by direction-dependent phase modulations.
We show that the generated electric field leads to Bloch oscillations
and the revival of the state after a certain number of steps dependent
on the field strength. We measure the probability of the revival and
demonstrate good agreement between the observed values and the theoretically
predicted results. Furthermore, by applying a nonuniform electric
field, we show the possibility of waveguiding photons. Ultimately,
our results open up new opportunities for manipulating the propagation
of photons with potential applications in photonic quantum simulations.
\end{abstract}
\maketitle
Photons are promising candidates for implementing quantum simulations
due to their wave characteristics that exhibit strong interference
effects. Recently, numerous complicated quantum simulations have been
performed using photonic systems by molding the flow of light in real
space \cite{Sparrow2018,Ma2011,Lanyon2010,Aspuru-Guzik2012}. However,
these systems are extremely difficult to scale and reconfigure. Synthetic
dimensions provide a promising alternative approach for photonic quantum
simulations in a scalable and resource efficient way without requiring
complex photonic circuits \cite{Navarrete-Benlloch2007,Bouwmeester1999,Bell2017,Qin2018,Dutt2019,Regensburger2012,Lustig2019,Lin2016,Lin2018,Yuan2018,Goyal2013,Cardano2015,Cardano2016,Cardano2017}.
One powerful technique to implement a synthetic space is through time-multiplexing
\cite{Schreiber2010,Schreiber2011,Schreiber2012,Nitsche2016,Barkhofen2017,Chen2018},
which can scale to a higher number of dimensions efficiently. But
significant challenges remain to fully control the evolution of photons
in synthetic spaces. 

One important challenge is that photons do not directly interact with
electromagnetic fields because of their lack of charge. This limitation
makes it difficult for photons to simulate the complex dynamical behavior
of electrons or atoms. But significant progress in the past decade
has led to the development of techniques to engineer magnetic and
electric fields for photons \cite{Hafezi2011,Mittal2014,Rechtsman2013,Fang2012}.
In particular, realization of synthetic magnetic fields has led to
the exploration of topological physics in photonic systems \cite{Ozawa2019},
and measurement of associated topological invariants \cite{Hu2015,Cardano2017,derrico2018twodimensional}.
Similarly, various ways to engineer electric fields have been reported
in real dimensional photonic circuits \cite{Bromberg2010,Corrielli2013,Witthaut2004}.
One outcome of applying a constant electric field in periodic systems
is the generation of state revivals known as Bloch oscillations, originally
predicted in electronic systems \cite{Bloch1929,Haller2010,BenDahan1996}.
Bloch oscillations have been observed in photonic systems such as
coupled optical waveguide arrays \cite{Pertsch1999,Morandotti1999,Lenz1999,Trompeter2006,Longhi2006,Dreisow2009,Levy2010}
and one-dimensional quantum walks of photons \cite{Xue_2015,Cedzich2016}.
Bloch oscillations in one dimension have also been explored using
frequency as a synthetic dimension \cite{Peschel2008,Bersch2009,Yuan2016v2}.
However, the extension of electric fields to two-dimensional synthetic
spaces has not yet been explored. 

Here we demonstrate an electric field for photons in a two-dimensional
synthetic space. We use time-multiplexing as a versatile platform
to create the synthetic space and a time-varying gauge field to create
the electric field. Under the application of a constant electric field,
we show that photons return to the original state after a certain
number of steps, thus demonstrating Bloch oscillations. Furthermore,
by generating a spatially nonuniform electric field, we realize a
synthetic quantum well, which guides photons without the use of a
bandgap. 

\begin{figure}
\includegraphics{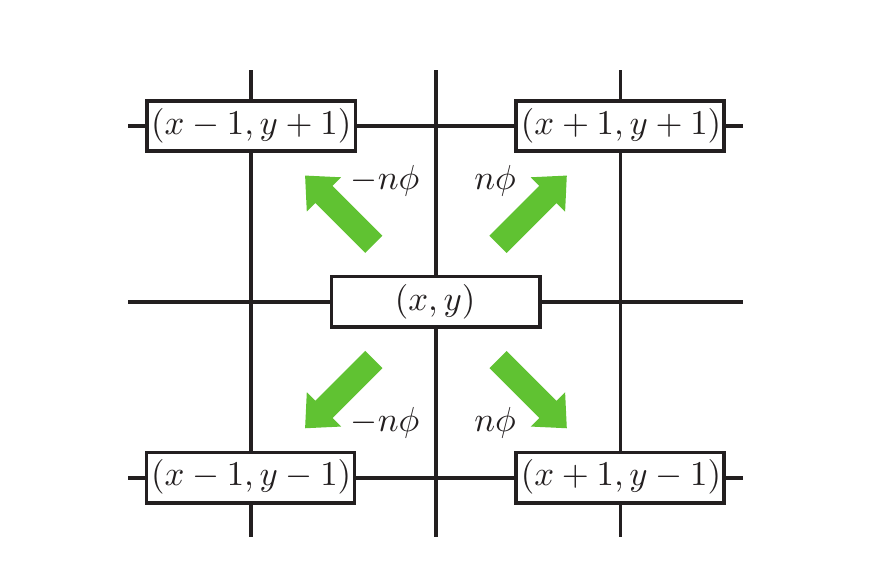}

\caption{\label{fig:Schematic}Applied phase modulation scheme for the nth
step of the quantum walk.}
\end{figure}

In our time-multiplexed photonic quantum walk, the quantum walker
state space is mapped into time delays of optical pulses. The experimental
setup is a closed-loop fiber architecture composed of two beam splitters
with their ports connected to fibers of different lengths mapping
the $\pm x$ and $\pm y$ directions to different time delays. Full
details of the experimental setup are explained in the Supplemental
Material \cite{supplem}. One complete propagation of an optical pulse
around the loop is equivalent to the hopping of the walker to one
of the four possible corners in the synthetic space (Fig. \ref{fig:Schematic}).
Semiconductor optical amplifiers as well as polarization controllers
are used in the setup to compensate for the losses and polarization
changes that the optical pulses experience in each round trip, respectively.
The quantum walk distribution at each time step is studied via two
photodetectors analyzing two channels that we refer to as the up and
down channels (See Fig. S1). A single incident laser pulse that is
injected into the up channel initializes the quantum walk evolution
from the origin in the synthetic space.

In this setup, we use electro-optic modulators that can introduce
desired phase shifts to pulses moving to the right or left directions
to generate the synthetic gauge field. Specifically, here we implement
a linearly time-varying gauge field ($\overrightarrow{A}=-\mathcal{E}t\hat{x}$,
in which $t$ denotes the time step and $\hat{x}$ is the unit vector
in the $x$ direction) that leads to the generation of a constant
electric field ($\overrightarrow{\mathcal{E}}=-\partial\overrightarrow{A}/\partial t$).
To implement this gauge field, a phase modulation needs to be applied
that varies with the time step. Figure \ref{fig:Schematic} depicts
the synthetic lattice with the required phase modulation criteria
describing the amount of phase that the walker accumulates in hopping
to the four possible corners at time step $n$. 

This method of generating an electric field is distinguished from
previously used approaches in discrete-time quantum walks, which have
been based on position-dependent but time-independent gauge fields
\cite{Cedzich2013,Genske2013,Bru2016}. In these approaches, an effective
linear electric potential $V=-\mathcal{E}x$ is implemented, which
leads to the generation of electric fields based on $\overrightarrow{\mathcal{E}}=-\nabla V$.
In order to create such a gauge field in discrete-time quantum walks,
the unitary operation in each time step must have an extra term relative
to the standard quantum walk evolution operator $U_{0}$ as $U_{\phi}=e^{i\phi x}U_{0}$.
In contrast, the current approach does not require any coordinate-dependent
unitary operation to generate an electric field. This is of particular
interest for time-multiplexed quantum walks, as it relaxes the need
for any variation of phase modulations during each time step. The
equivalence of these approaches can be understood in terms of a gauge
transformation \cite{Cedzich2019}. The similarity between the approach used and the conventional
coordinate-based method of implementing an electric field in a two-dimensional
quantum walk \cite{Bru2016} is described in the Supplemental Material
\cite{supplem}. We show that the total phase accumulated in some
sample closed loops that start from the origin and return to it in
both pictures are the same. In fact, this similarity holds for any
closed path starting from the origin and ending at it. 

\begin{figure}
\subfloat[\label{fig:phi90_timedep}]{}\subfloat[\label{fig:phi60_timedep}]{}\subfloat[\label{fig:phi45_timedep}]{}\includegraphics{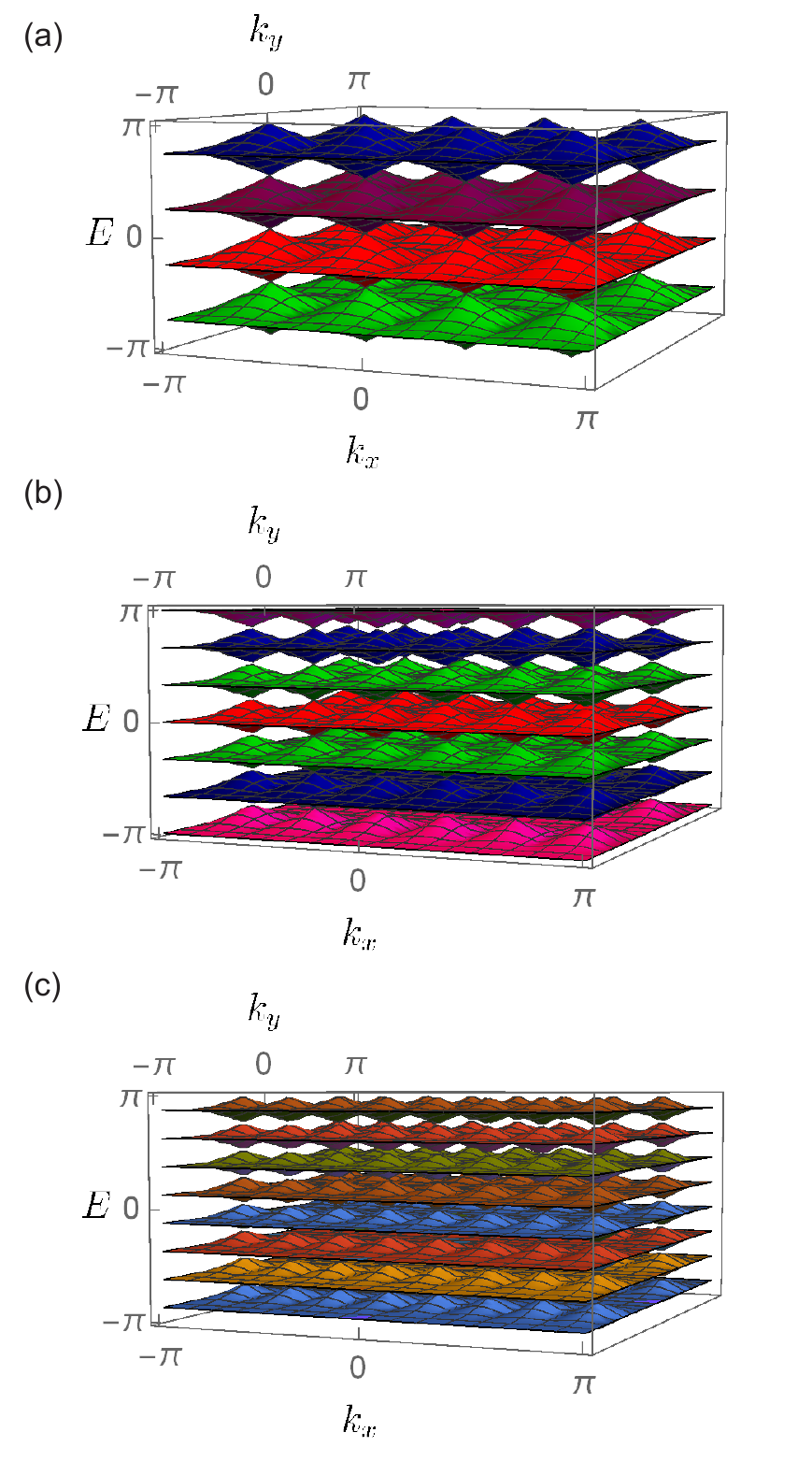}

\caption{\label{fig:banddiag_timedep}Pseudo energy band diagrams of the two-dimensional
quantum walk for different gauge field strengths: (a) $\phi=\pi/2$,
(b) $\phi=\pi/3$, and (c) $\phi=\pi/4$.}
\end{figure}

The application of the electric field in our two-dimensional discrete-time
quantum walk will lead to Bloch oscillations and the revival of the
quantum state. This can be intuitively explained through the band
diagram structure of the system. As we show in the Supplemental Material
\cite{supplem}, for a phase modulation with a fractional value of
$\phi$ as $\phi=2\pi/q$, the band structure has $2q$ bands. The
analytical expressions for the pseudo energy band structure under
such a phase modulation are given by:

\phantom{END}

\onecolumngrid 

\begin{figure}
\subfloat[\label{fig:evolution_exp}]{}\subfloat[\label{fig:evolution_th}]{}\includegraphics{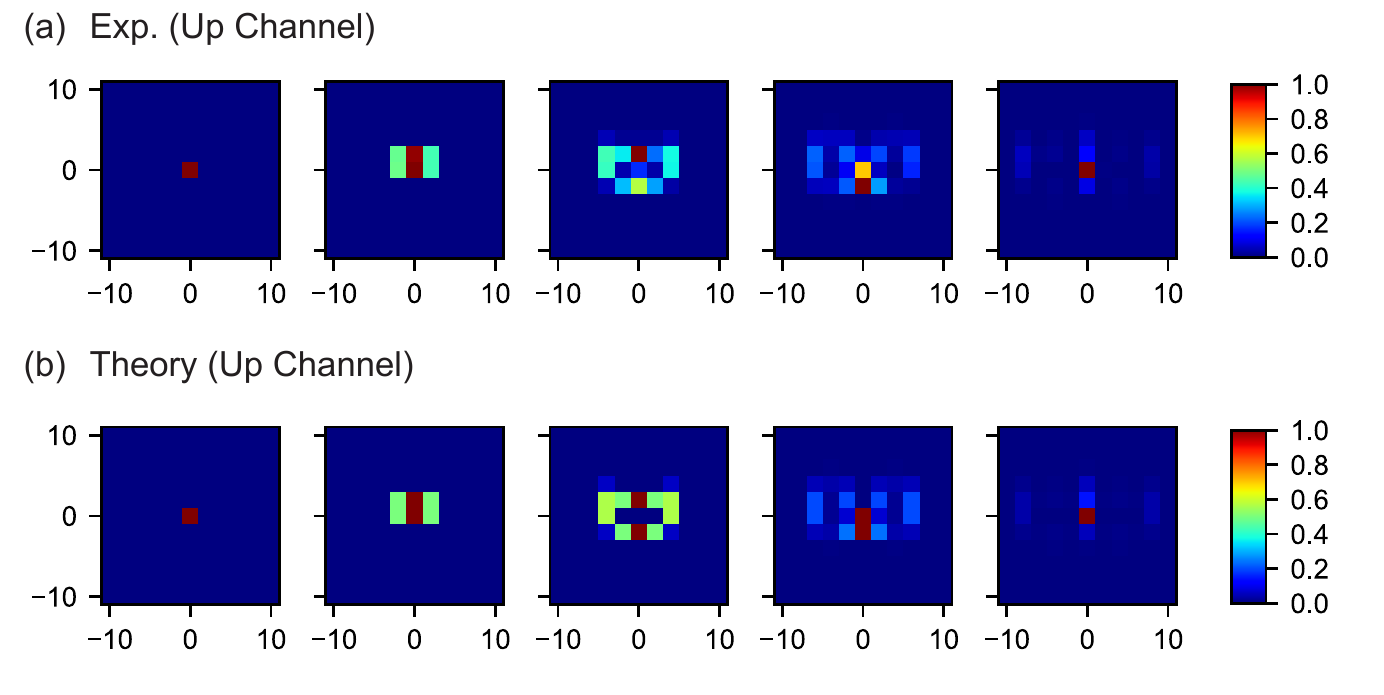}

\caption{\label{fig:evolution}(a) Experimental observations and (b) theoretical
predictions of the evolution of the quantum walk distribution under
time-dependent phase modulation for the case of $\phi=\pi/4$. The
columns from left to right show the distributions at time steps of
0, 2, 4, 6, and 8, respectively. In these plots, all the distributions
are normalized to their maximum.}
\end{figure}

\twocolumngrid

\begin{align*}
 & E_{n,\pm,k_{x},k_{y}}=\frac{2n\pi}{q}\pm\\
 & \frac{1}{q}\arccos\left(\cos\left(\frac{\pi q}{2}\right)\left(\sin^{q}k_{y}-1\right)-\cos\left(qk_{x}+\frac{\pi q}{2}\right)\sin^{q}k_{y}\right)
\end{align*}

where $k_{x}$ and $k_{y}$ are the momentum wave vectors in inverse
synthetic space and $n\in Z$. This expression for $q=1$ returns
to the form of $E_{\pm}=\pm\arccos\left(\sin\left(k_{x}\right)\sin\left(k_{y}\right)\right)$,
which represents the band structure for the quantum walk under no
effective applied gauge field \cite{Chalabi2019}. Figure \ref{fig:banddiag_timedep}
shows the band diagrams for three different values of the phase modulations.
By increasing $q$, the band structure becomes flatter and the corresponding
group velocities tend toward zero. As has been demonstrated in one
dimensional quantum walks, this will lead to the revival of the quantum
walk with high probability \cite{supplem}. This band flattening also
occurs in the two-dimensional Floquet quantum walks considered here,
which will lead to the return of the quantum walker toward the origin
after $q$ steps. In our system, the application of an electric field
in the $x$ direction will lead to the revival of the quantum walk
not only in the $x$ direction but also in the $y$ direction (see
Supplemental Material \cite{supplem}). 

To experimentally demonstrate Bloch oscillations in our 2D time-multiplexed
quantum walk caused by the applied electric field, we investigate
the evolution of the quantum walk distribution at different time steps.
In order to measure the quantum walk distribution, we measure the
power of the optical pulses received by the photodetectors at different
time delays for each time step. Figure \ref{fig:evolution_exp} shows
the experimentally measured quantum walk distributions at different
time steps. This figure demonstrates state revival under the application
of the time-varying gauge field due to the uniform electric field
generated. The phase strength of the applied electric field in this
case is $\phi=\pi/4$. As this figure shows, after 8 steps, the quantum
walker returns to the origin with high probability. The experimental
results are in good agreement with the corresponding theoretical predictions
shown in Fig. \ref{fig:evolution_th}. 

\begin{figure}
\begin{centering}
\includegraphics{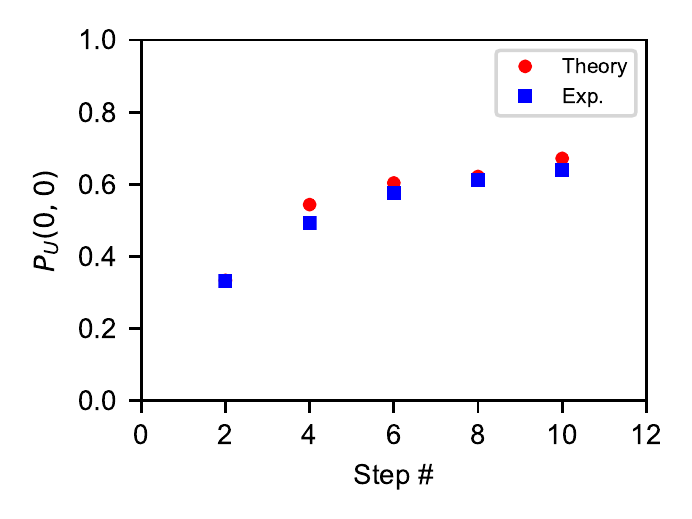}
\par\end{centering}
\caption{\label{fig:PU}Probability of revival: Experimental (blue solid squares)
and theoretical (red solid circles) probabilities of the walker returning
to the origin with respect to the required number of quantum walker
steps. The error bars in the measurements are smaller than the size
of the plotted data points.}
\end{figure}

To quantify the effect of the gauge field on the revival of the quantum
state, we measure the probability of the walker returning to the origin
(revival probability $P_{U}\left(0,0\right)$) as a function of the
number of steps taken (Fig. \ref{fig:PU}). We measure this probability
for different time-varying phase modulations after the appropriate
number of time steps ($2\pi/\phi$ steps) needed for the revival to
happen. As shown in Fig. \ref{fig:PU}, the revival probability increases
with increasing number of steps. Additionally, this figure indicates
that the experimental results are in good agreement with the theoretical
predictions. In the Supplemental Material \cite{supplem}, we demonstrate
that by decreasing the phase modulation, $\phi$, the revival probability
increases and tends toward unity. Using these measured probability
distributions, we can also calculate the statistical averages of the
distribution at different time steps. Specifically, we measured and
analyzed the quadratic means as well as the norm ones of the $x$
and $y$ coordinates at different time steps (Figs. S2 and S3). As
these results show, the quadratic means as well as the norm ones of
$x$ and $y$ tend toward the local minimum values after $2\pi/\phi$
steps. The variational behaviors of these quantities with the time
step confirm the revival effect in both the $x$ and $y$ directions
and are in good agreement with the theoretically predicted results. 

\begin{figure}
\subfloat[\label{fig:inhom_field_schem}]{}\subfloat[\label{fig:inhom_field_evolution_exp}]{}\subfloat[\label{fig:inhom_field_evolution_th}]{}\subfloat[\label{fig:inhom_field_qm}]{}\includegraphics{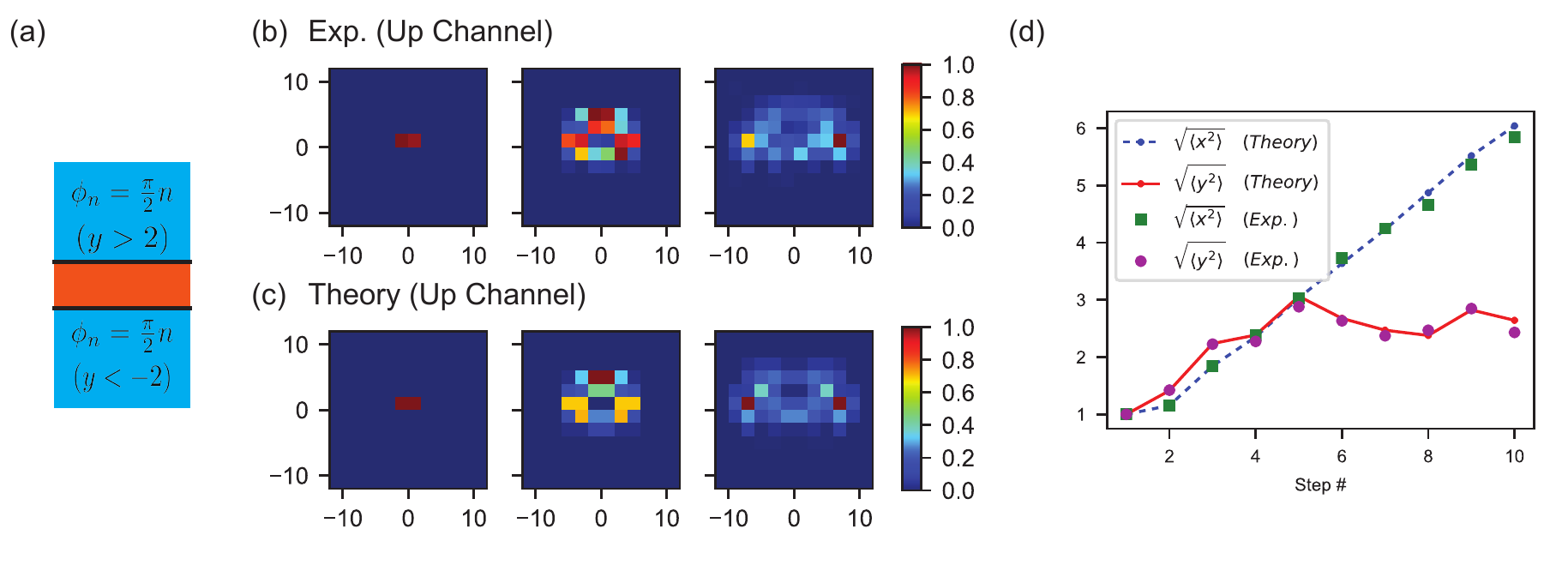}

\caption{\label{fig:inhom_field}Confinement of the quantum walker through
the application of a discontinuous electric field. (a) Schematic describing
the phase modulation pattern in the synthetic space that leads to
a zero electric field for $-2<y<2$ and a non-zero electric field
outside this range (b) Experimental observations and (c) theoretical
predictions of the evolution of the quantum walk distribution under
the discontinuous electric field. The columns from left to right show
the distributions at time steps of 1, 5, and 9, respectively. All
the distributions are normalized to their maximum. (d) Experimentally
measured and theoretically predicted quadratic means of x and y as
a function of the time step. The error bars in the measurements are
smaller than the size of the plotted data points.}
\end{figure}

We are not restricted to implementing only uniform electric fields
in our two-dimensional quantum walk. Spatially discontinuous electric
fields can also be created by simply controlling the modulation pattern
of the phase modulators. This provides the possibility to perform
waveguiding in the synthetic space using gauge fields. Figure \ref{fig:inhom_field_schem}
shows an example of a discontinuous electric field in a two-dimensional
synthetic space. In this configuration, the electric field is zero
in the $-2<y<2$ region and non-zero outside this range. Due to the
boundary created by the discontinuous gauge fields, the quantum random
walk evolution is mainly confined to the region where the electric
field is equal to zero. Figure \ref{fig:inhom_field_evolution_exp}
presents the experimentally measured quantum walk distributions at
different time steps showing the trapping caused by the existence
of the boundaries in the electric field pattern. The experimentally
measured results are in good agreement with the theoretical predictions
shown in Fig. \ref{fig:inhom_field_evolution_th}. Additionally, more
quantitative agreement can be inferred based on Fig. \ref{fig:inhom_field_qm}
depicting the variations of the quadratic means as functions of the
time step. This figure clearly shows that the presence of boundaries
in the field pattern has led to the confinement of the quantum walk
in the y ribbon. We note that this confinement is not induced by a
bandgap and is therefore physically distinct from the confinement
in conventional crystal heterostructures. None of the three regions
shown in Fig. \ref{fig:inhom_field} supports a bandgap, thus demonstrating
the confinement is directly induced by the gauge field itself. These
results demonstrate the possibility of using a nonuniform electric
field in order to guide the path of the quantum walk in a desired
fashion. 

In conclusion, we studied the time evolution of a quantum random walk
under a time-varying gauge field in a two-dimensional synthetic space.
Using a linearly time-varying phase modulation, an electric field
acting on photonic quantum walkers can be created. Our findings demonstrate
that under the influence of such an electric field a complete revival
caused by Bloch oscillations happens in two-dimensional quantum walks.
This revival becomes more accurate as we increase the number of steps.
Moreover, we demonstrated that a discontinuous electric field could
impose confinement on the evolution of a quantum random walk, even
when there is no bandgap. While we demonstrated the Bloch oscillation
for a quantum walk initiated with classical coherent laser pulses,
the same physics holds at the single photon level. The obtained results
can be extended to the investigation of the effect of dynamic localization
\cite{Dunlap1986,Lenz2003,Longhi2006v2,Iyer2007,Joushaghani2009,Joushaghani2012,Yuan2015}.
Our demonstration of an electric field for photons in time-multiplexed
synthetic lattices will have potential applications in photonic quantum
simulations, for example, multi-photon interference and Boson sampling
in the time-domain \cite{Orre2019}, and the realization of photonic
lattices with strong nonlinearities mediated via artificial atoms
like quantum dots \cite{Pichler2017,Chalabi2018}.

This work was supported by the Air Force Office of Scientific Research-Multidisciplinary
University Research Initiative (Grant No. FA9550-16-1-0323), the Physics
Frontier Center at the Joint Quantum Institute, the National Science
Foundation (Grant No. PHYS. 1415458 as well as PHY-1430094), and the
Center for Distributed Quantum Information. The authors would also
like to acknowledge support from the U.S. Department of Defense. Moreover,
the authors acknowledge support from the Laboratory for Telecommunication
Sciences.

\phantom{END}

\onecolumngrid 

\twocolumngrid

\phantom{END}

\bibliographystyle{apsrev4-1}
\bibliography{refers_arXiv}

\begin{thebibliography}{65}%
\makeatletter
\providecommand \@ifxundefined [1]{%
 \@ifx{#1\undefined}
}%
\providecommand \@ifnum [1]{%
 \ifnum #1\expandafter \@firstoftwo
 \else \expandafter \@secondoftwo
 \fi
}%
\providecommand \@ifx [1]{%
 \ifx #1\expandafter \@firstoftwo
 \else \expandafter \@secondoftwo
 \fi
}%
\providecommand \natexlab [1]{#1}%
\providecommand \enquote  [1]{``#1''}%
\providecommand \bibnamefont  [1]{#1}%
\providecommand \bibfnamefont [1]{#1}%
\providecommand \citenamefont [1]{#1}%
\providecommand \href@noop [0]{\@secondoftwo}%
\providecommand \href [0]{\begingroup \@sanitize@url \@href}%
\providecommand \@href[1]{\@@startlink{#1}\@@href}%
\providecommand \@@href[1]{\endgroup#1\@@endlink}%
\providecommand \@sanitize@url [0]{\catcode `\\12\catcode `\$12\catcode
  `\&12\catcode `\#12\catcode `\^12\catcode `\_12\catcode `\%12\relax}%
\providecommand \@@startlink[1]{}%
\providecommand \@@endlink[0]{}%
\providecommand \url  [0]{\begingroup\@sanitize@url \@url }%
\providecommand \@url [1]{\endgroup\@href {#1}{\urlprefix }}%
\providecommand \urlprefix  [0]{URL }%
\providecommand \Eprint [0]{\href }%
\providecommand \doibase [0]{http://dx.doi.org/}%
\providecommand \selectlanguage [0]{\@gobble}%
\providecommand \bibinfo  [0]{\@secondoftwo}%
\providecommand \bibfield  [0]{\@secondoftwo}%
\providecommand \translation [1]{[#1]}%
\providecommand \BibitemOpen [0]{}%
\providecommand \bibitemStop [0]{}%
\providecommand \bibitemNoStop [0]{.\EOS\space}%
\providecommand \EOS [0]{\spacefactor3000\relax}%
\providecommand \BibitemShut  [1]{\csname bibitem#1\endcsname}%
\let\auto@bib@innerbib\@empty
%</preamble>
\bibitem [{\citenamefont {Sparrow}\ \emph {et~al.}(2018)\citenamefont
  {Sparrow}, \citenamefont {Mart{\'{i}}n-L{\'{o}}pez}, \citenamefont
  {Maraviglia}, \citenamefont {Neville}, \citenamefont {Harrold}, \citenamefont
  {Carolan}, \citenamefont {Joglekar}, \citenamefont {Hashimoto}, \citenamefont
  {Matsuda}, \citenamefont {O'Brien}, \citenamefont {Tew},\ and\ \citenamefont
  {Laing}}]{Sparrow2018}%
  \BibitemOpen
  \bibfield  {author} {\bibinfo {author} {\bibfnamefont {C.}~\bibnamefont
  {Sparrow}}, \bibinfo {author} {\bibfnamefont {E.}~\bibnamefont
  {Mart{\'{i}}n-L{\'{o}}pez}}, \bibinfo {author} {\bibfnamefont
  {N.}~\bibnamefont {Maraviglia}}, \bibinfo {author} {\bibfnamefont
  {A.}~\bibnamefont {Neville}}, \bibinfo {author} {\bibfnamefont
  {C.}~\bibnamefont {Harrold}}, \bibinfo {author} {\bibfnamefont
  {J.}~\bibnamefont {Carolan}}, \bibinfo {author} {\bibfnamefont {Y.~N.}\
  \bibnamefont {Joglekar}}, \bibinfo {author} {\bibfnamefont {T.}~\bibnamefont
  {Hashimoto}}, \bibinfo {author} {\bibfnamefont {N.}~\bibnamefont {Matsuda}},
  \bibinfo {author} {\bibfnamefont {J.~L.}\ \bibnamefont {O'Brien}}, \bibinfo
  {author} {\bibfnamefont {D.~P.}\ \bibnamefont {Tew}}, \ and\ \bibinfo
  {author} {\bibfnamefont {A.}~\bibnamefont {Laing}},\ }\href {\doibase
  10.1038/s41586-018-0152-9} {\bibfield  {journal} {\bibinfo  {journal}
  {Nature}\ }\textbf {\bibinfo {volume} {557}},\ \bibinfo {pages} {660}
  (\bibinfo {year} {2018})}\BibitemShut {NoStop}%
\bibitem [{\citenamefont {Ma}\ \emph {et~al.}(2011)\citenamefont {Ma},
  \citenamefont {Dakic}, \citenamefont {Naylor}, \citenamefont {Zeilinger},\
  and\ \citenamefont {Walther}}]{Ma2011}%
  \BibitemOpen
  \bibfield  {author} {\bibinfo {author} {\bibfnamefont {X.-s.}\ \bibnamefont
  {Ma}}, \bibinfo {author} {\bibfnamefont {B.}~\bibnamefont {Dakic}}, \bibinfo
  {author} {\bibfnamefont {W.}~\bibnamefont {Naylor}}, \bibinfo {author}
  {\bibfnamefont {A.}~\bibnamefont {Zeilinger}}, \ and\ \bibinfo {author}
  {\bibfnamefont {P.}~\bibnamefont {Walther}},\ }\href {\doibase
  10.1038/nphys1919} {\bibfield  {journal} {\bibinfo  {journal} {Nature
  Physics}\ }\textbf {\bibinfo {volume} {7}},\ \bibinfo {pages} {399} (\bibinfo
  {year} {2011})}\BibitemShut {NoStop}%
\bibitem [{\citenamefont {Lanyon}\ \emph {et~al.}(2010)\citenamefont {Lanyon},
  \citenamefont {Whitfield}, \citenamefont {Gillett}, \citenamefont {Goggin},
  \citenamefont {Almeida}, \citenamefont {Kassal}, \citenamefont {Biamonte},
  \citenamefont {Mohseni}, \citenamefont {Powell}, \citenamefont {Barbieri},
  \citenamefont {Aspuru-Guzik},\ and\ \citenamefont {White}}]{Lanyon2010}%
  \BibitemOpen
  \bibfield  {author} {\bibinfo {author} {\bibfnamefont {B.~P.}\ \bibnamefont
  {Lanyon}}, \bibinfo {author} {\bibfnamefont {J.~D.}\ \bibnamefont
  {Whitfield}}, \bibinfo {author} {\bibfnamefont {G.~G.}\ \bibnamefont
  {Gillett}}, \bibinfo {author} {\bibfnamefont {M.~E.}\ \bibnamefont {Goggin}},
  \bibinfo {author} {\bibfnamefont {M.~P.}\ \bibnamefont {Almeida}}, \bibinfo
  {author} {\bibfnamefont {I.}~\bibnamefont {Kassal}}, \bibinfo {author}
  {\bibfnamefont {J.~D.}\ \bibnamefont {Biamonte}}, \bibinfo {author}
  {\bibfnamefont {M.}~\bibnamefont {Mohseni}}, \bibinfo {author} {\bibfnamefont
  {B.~J.}\ \bibnamefont {Powell}}, \bibinfo {author} {\bibfnamefont
  {M.}~\bibnamefont {Barbieri}}, \bibinfo {author} {\bibfnamefont
  {A.}~\bibnamefont {Aspuru-Guzik}}, \ and\ \bibinfo {author} {\bibfnamefont
  {A.~G.}\ \bibnamefont {White}},\ }\href {\doibase 10.1038/nchem.483}
  {\bibfield  {journal} {\bibinfo  {journal} {Nature Chemistry}\ }\textbf
  {\bibinfo {volume} {2}},\ \bibinfo {pages} {106} (\bibinfo {year}
  {2010})}\BibitemShut {NoStop}%
\bibitem [{\citenamefont {Aspuru-Guzik}\ and\ \citenamefont
  {Walther}(2012)}]{Aspuru-Guzik2012}%
  \BibitemOpen
  \bibfield  {author} {\bibinfo {author} {\bibfnamefont {A.}~\bibnamefont
  {Aspuru-Guzik}}\ and\ \bibinfo {author} {\bibfnamefont {P.}~\bibnamefont
  {Walther}},\ }\href {\doibase 10.1038/nphys2253} {\bibfield  {journal}
  {\bibinfo  {journal} {Nature Physics}\ }\textbf {\bibinfo {volume} {8}},\
  \bibinfo {pages} {285} (\bibinfo {year} {2012})}\BibitemShut {NoStop}%
\bibitem [{\citenamefont {Navarrete-Benlloch}\ \emph
  {et~al.}(2007)\citenamefont {Navarrete-Benlloch}, \citenamefont
  {P{\'{e}}rez},\ and\ \citenamefont {Rold{\'{a}}n}}]{Navarrete-Benlloch2007}%
  \BibitemOpen
  \bibfield  {author} {\bibinfo {author} {\bibfnamefont {C.}~\bibnamefont
  {Navarrete-Benlloch}}, \bibinfo {author} {\bibfnamefont {A.}~\bibnamefont
  {P{\'{e}}rez}}, \ and\ \bibinfo {author} {\bibfnamefont {E.}~\bibnamefont
  {Rold{\'{a}}n}},\ }\href {\doibase 10.1103/PhysRevA.75.062333} {\bibfield
  {journal} {\bibinfo  {journal} {Physical Review A}\ }\textbf {\bibinfo
  {volume} {75}},\ \bibinfo {pages} {062333} (\bibinfo {year}
  {2007})}\BibitemShut {NoStop}%
\bibitem [{\citenamefont {Bouwmeester}\ \emph {et~al.}(1999)\citenamefont
  {Bouwmeester}, \citenamefont {Marzoli}, \citenamefont {Karman}, \citenamefont
  {Schleich},\ and\ \citenamefont {Woerdman}}]{Bouwmeester1999}%
  \BibitemOpen
  \bibfield  {author} {\bibinfo {author} {\bibfnamefont {D.}~\bibnamefont
  {Bouwmeester}}, \bibinfo {author} {\bibfnamefont {I.}~\bibnamefont
  {Marzoli}}, \bibinfo {author} {\bibfnamefont {G.~P.}\ \bibnamefont {Karman}},
  \bibinfo {author} {\bibfnamefont {W.}~\bibnamefont {Schleich}}, \ and\
  \bibinfo {author} {\bibfnamefont {J.~P.}\ \bibnamefont {Woerdman}},\ }\href
  {\doibase 10.1103/PhysRevA.61.013410} {\bibfield  {journal} {\bibinfo
  {journal} {Physical Review A}\ }\textbf {\bibinfo {volume} {61}},\ \bibinfo
  {pages} {013410} (\bibinfo {year} {1999})}\BibitemShut {NoStop}%
\bibitem [{\citenamefont {Bell}\ \emph {et~al.}(2017)\citenamefont {Bell},
  \citenamefont {Wang}, \citenamefont {Solntsev}, \citenamefont {Neshev},
  \citenamefont {Sukhorukov},\ and\ \citenamefont {Eggleton}}]{Bell2017}%
  \BibitemOpen
  \bibfield  {author} {\bibinfo {author} {\bibfnamefont {B.~A.}\ \bibnamefont
  {Bell}}, \bibinfo {author} {\bibfnamefont {K.}~\bibnamefont {Wang}}, \bibinfo
  {author} {\bibfnamefont {A.~S.}\ \bibnamefont {Solntsev}}, \bibinfo {author}
  {\bibfnamefont {D.~N.}\ \bibnamefont {Neshev}}, \bibinfo {author}
  {\bibfnamefont {A.~A.}\ \bibnamefont {Sukhorukov}}, \ and\ \bibinfo {author}
  {\bibfnamefont {B.~J.}\ \bibnamefont {Eggleton}},\ }\href {\doibase
  10.1364/OPTICA.4.001433} {\bibfield  {journal} {\bibinfo  {journal} {Optica}\
  }\textbf {\bibinfo {volume} {4}},\ \bibinfo {pages} {1433} (\bibinfo {year}
  {2017})}\BibitemShut {NoStop}%
\bibitem [{\citenamefont {Qin}\ \emph {et~al.}(2018)\citenamefont {Qin},
  \citenamefont {Zhou}, \citenamefont {Peng}, \citenamefont {Sounas},
  \citenamefont {Zhu}, \citenamefont {Wang}, \citenamefont {Dong},
  \citenamefont {Zhang}, \citenamefont {Al{\`{u}}},\ and\ \citenamefont
  {Lu}}]{Qin2018}%
  \BibitemOpen
  \bibfield  {author} {\bibinfo {author} {\bibfnamefont {C.}~\bibnamefont
  {Qin}}, \bibinfo {author} {\bibfnamefont {F.}~\bibnamefont {Zhou}}, \bibinfo
  {author} {\bibfnamefont {Y.}~\bibnamefont {Peng}}, \bibinfo {author}
  {\bibfnamefont {D.}~\bibnamefont {Sounas}}, \bibinfo {author} {\bibfnamefont
  {X.}~\bibnamefont {Zhu}}, \bibinfo {author} {\bibfnamefont {B.}~\bibnamefont
  {Wang}}, \bibinfo {author} {\bibfnamefont {J.}~\bibnamefont {Dong}}, \bibinfo
  {author} {\bibfnamefont {X.}~\bibnamefont {Zhang}}, \bibinfo {author}
  {\bibfnamefont {A.}~\bibnamefont {Al{\`{u}}}}, \ and\ \bibinfo {author}
  {\bibfnamefont {P.}~\bibnamefont {Lu}},\ }\href {\doibase
  10.1103/PhysRevLett.120.133901} {\bibfield  {journal} {\bibinfo  {journal}
  {Physical Review Letters}\ }\textbf {\bibinfo {volume} {120}},\ \bibinfo
  {pages} {133901} (\bibinfo {year} {2018})}\BibitemShut {NoStop}%
\bibitem [{\citenamefont {Dutt}\ \emph {et~al.}(2019)\citenamefont {Dutt},
  \citenamefont {Minkov}, \citenamefont {Lin}, \citenamefont {Yuan},
  \citenamefont {Miller},\ and\ \citenamefont {Fan}}]{Dutt2019}%
  \BibitemOpen
  \bibfield  {author} {\bibinfo {author} {\bibfnamefont {A.}~\bibnamefont
  {Dutt}}, \bibinfo {author} {\bibfnamefont {M.}~\bibnamefont {Minkov}},
  \bibinfo {author} {\bibfnamefont {Q.}~\bibnamefont {Lin}}, \bibinfo {author}
  {\bibfnamefont {L.}~\bibnamefont {Yuan}}, \bibinfo {author} {\bibfnamefont
  {D.~A.~B.}\ \bibnamefont {Miller}}, \ and\ \bibinfo {author} {\bibfnamefont
  {S.}~\bibnamefont {Fan}},\ }\href {\doibase 10.1038/s41467-019-11117-9}
  {\bibfield  {journal} {\bibinfo  {journal} {Nature Communications}\ }\textbf
  {\bibinfo {volume} {10}},\ \bibinfo {pages} {3122} (\bibinfo {year}
  {2019})},\ \Eprint {http://arxiv.org/abs/1903.07842} {arXiv:1903.07842}
  \BibitemShut {NoStop}%
\bibitem [{\citenamefont {Regensburger}\ \emph {et~al.}(2012)\citenamefont
  {Regensburger}, \citenamefont {Bersch}, \citenamefont {Miri}, \citenamefont
  {Onishchukov}, \citenamefont {Christodoulides},\ and\ \citenamefont
  {Peschel}}]{Regensburger2012}%
  \BibitemOpen
  \bibfield  {author} {\bibinfo {author} {\bibfnamefont {A.}~\bibnamefont
  {Regensburger}}, \bibinfo {author} {\bibfnamefont {C.}~\bibnamefont
  {Bersch}}, \bibinfo {author} {\bibfnamefont {M.-A.}\ \bibnamefont {Miri}},
  \bibinfo {author} {\bibfnamefont {G.}~\bibnamefont {Onishchukov}}, \bibinfo
  {author} {\bibfnamefont {D.~N.}\ \bibnamefont {Christodoulides}}, \ and\
  \bibinfo {author} {\bibfnamefont {U.}~\bibnamefont {Peschel}},\ }\href
  {\doibase 10.1038/nature11298} {\bibfield  {journal} {\bibinfo  {journal}
  {Nature}\ }\textbf {\bibinfo {volume} {488}},\ \bibinfo {pages} {167}
  (\bibinfo {year} {2012})}\BibitemShut {NoStop}%
\bibitem [{\citenamefont {Lustig}\ \emph {et~al.}(2019)\citenamefont {Lustig},
  \citenamefont {Weimann}, \citenamefont {Plotnik}, \citenamefont {Lumer},
  \citenamefont {Bandres}, \citenamefont {Szameit},\ and\ \citenamefont
  {Segev}}]{Lustig2019}%
  \BibitemOpen
  \bibfield  {author} {\bibinfo {author} {\bibfnamefont {E.}~\bibnamefont
  {Lustig}}, \bibinfo {author} {\bibfnamefont {S.}~\bibnamefont {Weimann}},
  \bibinfo {author} {\bibfnamefont {Y.}~\bibnamefont {Plotnik}}, \bibinfo
  {author} {\bibfnamefont {Y.}~\bibnamefont {Lumer}}, \bibinfo {author}
  {\bibfnamefont {M.~A.}\ \bibnamefont {Bandres}}, \bibinfo {author}
  {\bibfnamefont {A.}~\bibnamefont {Szameit}}, \ and\ \bibinfo {author}
  {\bibfnamefont {M.}~\bibnamefont {Segev}},\ }\href {\doibase
  10.1038/s41586-019-0943-7} {\bibfield  {journal} {\bibinfo  {journal}
  {Nature}\ }\textbf {\bibinfo {volume} {567}},\ \bibinfo {pages} {356}
  (\bibinfo {year} {2019})}\BibitemShut {NoStop}%
\bibitem [{\citenamefont {Lin}\ \emph {et~al.}(2016)\citenamefont {Lin},
  \citenamefont {Xiao}, \citenamefont {Yuan},\ and\ \citenamefont
  {Fan}}]{Lin2016}%
  \BibitemOpen
  \bibfield  {author} {\bibinfo {author} {\bibfnamefont {Q.}~\bibnamefont
  {Lin}}, \bibinfo {author} {\bibfnamefont {M.}~\bibnamefont {Xiao}}, \bibinfo
  {author} {\bibfnamefont {L.}~\bibnamefont {Yuan}}, \ and\ \bibinfo {author}
  {\bibfnamefont {S.}~\bibnamefont {Fan}},\ }\href {\doibase
  10.1038/ncomms13731} {\bibfield  {journal} {\bibinfo  {journal} {Nature
  Communications}\ }\textbf {\bibinfo {volume} {7}},\ \bibinfo {pages} {13731}
  (\bibinfo {year} {2016})}\BibitemShut {NoStop}%
\bibitem [{\citenamefont {Lin}\ \emph {et~al.}(2018)\citenamefont {Lin},
  \citenamefont {Sun}, \citenamefont {Xiao}, \citenamefont {Zhang},\ and\
  \citenamefont {Fan}}]{Lin2018}%
  \BibitemOpen
  \bibfield  {author} {\bibinfo {author} {\bibfnamefont {Q.}~\bibnamefont
  {Lin}}, \bibinfo {author} {\bibfnamefont {X.-Q.}\ \bibnamefont {Sun}},
  \bibinfo {author} {\bibfnamefont {M.}~\bibnamefont {Xiao}}, \bibinfo {author}
  {\bibfnamefont {S.-C.}\ \bibnamefont {Zhang}}, \ and\ \bibinfo {author}
  {\bibfnamefont {S.}~\bibnamefont {Fan}},\ }\href {\doibase
  10.1126/sciadv.aat2774} {\bibfield  {journal} {\bibinfo  {journal} {Science
  Advances}\ }\textbf {\bibinfo {volume} {4}},\ \bibinfo {pages} {eaat2774}
  (\bibinfo {year} {2018})}\BibitemShut {NoStop}%
\bibitem [{\citenamefont {Yuan}\ \emph {et~al.}(2018)\citenamefont {Yuan},
  \citenamefont {Xiao}, \citenamefont {Lin},\ and\ \citenamefont
  {Fan}}]{Yuan2018}%
  \BibitemOpen
  \bibfield  {author} {\bibinfo {author} {\bibfnamefont {L.}~\bibnamefont
  {Yuan}}, \bibinfo {author} {\bibfnamefont {M.}~\bibnamefont {Xiao}}, \bibinfo
  {author} {\bibfnamefont {Q.}~\bibnamefont {Lin}}, \ and\ \bibinfo {author}
  {\bibfnamefont {S.}~\bibnamefont {Fan}},\ }\href {\doibase
  10.1103/PhysRevB.97.104105} {\bibfield  {journal} {\bibinfo  {journal}
  {Physical Review B}\ }\textbf {\bibinfo {volume} {97}},\ \bibinfo {pages}
  {104105} (\bibinfo {year} {2018})},\ \Eprint
  {http://arxiv.org/abs/1710.01373} {arXiv:1710.01373} \BibitemShut {NoStop}%
\bibitem [{\citenamefont {Goyal}\ \emph {et~al.}(2013)\citenamefont {Goyal},
  \citenamefont {Roux}, \citenamefont {Forbes},\ and\ \citenamefont
  {Konrad}}]{Goyal2013}%
  \BibitemOpen
  \bibfield  {author} {\bibinfo {author} {\bibfnamefont {S.~K.}\ \bibnamefont
  {Goyal}}, \bibinfo {author} {\bibfnamefont {F.~S.}\ \bibnamefont {Roux}},
  \bibinfo {author} {\bibfnamefont {A.}~\bibnamefont {Forbes}}, \ and\ \bibinfo
  {author} {\bibfnamefont {T.}~\bibnamefont {Konrad}},\ }\href {\doibase
  10.1103/PhysRevLett.110.263602} {\bibfield  {journal} {\bibinfo  {journal}
  {Physical Review Letters}\ }\textbf {\bibinfo {volume} {110}},\ \bibinfo
  {pages} {263602} (\bibinfo {year} {2013})}\BibitemShut {NoStop}%
\bibitem [{\citenamefont {Cardano}\ \emph {et~al.}(2015)\citenamefont
  {Cardano}, \citenamefont {Massa}, \citenamefont {Qassim}, \citenamefont
  {Karimi}, \citenamefont {Slussarenko}, \citenamefont {Paparo}, \citenamefont
  {de~Lisio}, \citenamefont {Sciarrino}, \citenamefont {Santamato},
  \citenamefont {Boyd},\ and\ \citenamefont {Marrucci}}]{Cardano2015}%
  \BibitemOpen
  \bibfield  {author} {\bibinfo {author} {\bibfnamefont {F.}~\bibnamefont
  {Cardano}}, \bibinfo {author} {\bibfnamefont {F.}~\bibnamefont {Massa}},
  \bibinfo {author} {\bibfnamefont {H.}~\bibnamefont {Qassim}}, \bibinfo
  {author} {\bibfnamefont {E.}~\bibnamefont {Karimi}}, \bibinfo {author}
  {\bibfnamefont {S.}~\bibnamefont {Slussarenko}}, \bibinfo {author}
  {\bibfnamefont {D.}~\bibnamefont {Paparo}}, \bibinfo {author} {\bibfnamefont
  {C.}~\bibnamefont {de~Lisio}}, \bibinfo {author} {\bibfnamefont
  {F.}~\bibnamefont {Sciarrino}}, \bibinfo {author} {\bibfnamefont
  {E.}~\bibnamefont {Santamato}}, \bibinfo {author} {\bibfnamefont {R.~W.}\
  \bibnamefont {Boyd}}, \ and\ \bibinfo {author} {\bibfnamefont
  {L.}~\bibnamefont {Marrucci}},\ }\href {\doibase 10.1126/sciadv.1500087}
  {\bibfield  {journal} {\bibinfo  {journal} {Science Advances}\ }\textbf
  {\bibinfo {volume} {1}},\ \bibinfo {pages} {e1500087} (\bibinfo {year}
  {2015})}\BibitemShut {NoStop}%
\bibitem [{\citenamefont {Cardano}\ \emph {et~al.}(2016)\citenamefont
  {Cardano}, \citenamefont {Maffei}, \citenamefont {Massa}, \citenamefont
  {Piccirillo}, \citenamefont {de~Lisio}, \citenamefont {{De Filippis}},
  \citenamefont {Cataudella}, \citenamefont {Santamato},\ and\ \citenamefont
  {Marrucci}}]{Cardano2016}%
  \BibitemOpen
  \bibfield  {author} {\bibinfo {author} {\bibfnamefont {F.}~\bibnamefont
  {Cardano}}, \bibinfo {author} {\bibfnamefont {M.}~\bibnamefont {Maffei}},
  \bibinfo {author} {\bibfnamefont {F.}~\bibnamefont {Massa}}, \bibinfo
  {author} {\bibfnamefont {B.}~\bibnamefont {Piccirillo}}, \bibinfo {author}
  {\bibfnamefont {C.}~\bibnamefont {de~Lisio}}, \bibinfo {author}
  {\bibfnamefont {G.}~\bibnamefont {{De Filippis}}}, \bibinfo {author}
  {\bibfnamefont {V.}~\bibnamefont {Cataudella}}, \bibinfo {author}
  {\bibfnamefont {E.}~\bibnamefont {Santamato}}, \ and\ \bibinfo {author}
  {\bibfnamefont {L.}~\bibnamefont {Marrucci}},\ }\href {\doibase
  10.1038/ncomms11439} {\bibfield  {journal} {\bibinfo  {journal} {Nature
  Communications}\ }\textbf {\bibinfo {volume} {7}},\ \bibinfo {pages} {11439}
  (\bibinfo {year} {2016})}\BibitemShut {NoStop}%
\bibitem [{\citenamefont {Cardano}\ \emph {et~al.}(2017)\citenamefont
  {Cardano}, \citenamefont {D'Errico}, \citenamefont {Dauphin}, \citenamefont
  {Maffei}, \citenamefont {Piccirillo}, \citenamefont {de~Lisio}, \citenamefont
  {{De Filippis}}, \citenamefont {Cataudella}, \citenamefont {Santamato},
  \citenamefont {Marrucci}, \citenamefont {Lewenstein},\ and\ \citenamefont
  {Massignan}}]{Cardano2017}%
  \BibitemOpen
  \bibfield  {author} {\bibinfo {author} {\bibfnamefont {F.}~\bibnamefont
  {Cardano}}, \bibinfo {author} {\bibfnamefont {A.}~\bibnamefont {D'Errico}},
  \bibinfo {author} {\bibfnamefont {A.}~\bibnamefont {Dauphin}}, \bibinfo
  {author} {\bibfnamefont {M.}~\bibnamefont {Maffei}}, \bibinfo {author}
  {\bibfnamefont {B.}~\bibnamefont {Piccirillo}}, \bibinfo {author}
  {\bibfnamefont {C.}~\bibnamefont {de~Lisio}}, \bibinfo {author}
  {\bibfnamefont {G.}~\bibnamefont {{De Filippis}}}, \bibinfo {author}
  {\bibfnamefont {V.}~\bibnamefont {Cataudella}}, \bibinfo {author}
  {\bibfnamefont {E.}~\bibnamefont {Santamato}}, \bibinfo {author}
  {\bibfnamefont {L.}~\bibnamefont {Marrucci}}, \bibinfo {author}
  {\bibfnamefont {M.}~\bibnamefont {Lewenstein}}, \ and\ \bibinfo {author}
  {\bibfnamefont {P.}~\bibnamefont {Massignan}},\ }\href {\doibase
  10.1038/ncomms15516} {\bibfield  {journal} {\bibinfo  {journal} {Nature
  Communications}\ }\textbf {\bibinfo {volume} {8}},\ \bibinfo {pages} {15516}
  (\bibinfo {year} {2017})}\BibitemShut {NoStop}%
\bibitem [{\citenamefont {Schreiber}\ \emph {et~al.}(2010)\citenamefont
  {Schreiber}, \citenamefont {Cassemiro}, \citenamefont {Poto{\v{c}}ek},
  \citenamefont {G{\'{a}}bris}, \citenamefont {Mosley}, \citenamefont
  {Andersson}, \citenamefont {Jex},\ and\ \citenamefont
  {Silberhorn}}]{Schreiber2010}%
  \BibitemOpen
  \bibfield  {author} {\bibinfo {author} {\bibfnamefont {A.}~\bibnamefont
  {Schreiber}}, \bibinfo {author} {\bibfnamefont {K.~N.}\ \bibnamefont
  {Cassemiro}}, \bibinfo {author} {\bibfnamefont {V.}~\bibnamefont
  {Poto{\v{c}}ek}}, \bibinfo {author} {\bibfnamefont {A.}~\bibnamefont
  {G{\'{a}}bris}}, \bibinfo {author} {\bibfnamefont {P.~J.}\ \bibnamefont
  {Mosley}}, \bibinfo {author} {\bibfnamefont {E.}~\bibnamefont {Andersson}},
  \bibinfo {author} {\bibfnamefont {I.}~\bibnamefont {Jex}}, \ and\ \bibinfo
  {author} {\bibfnamefont {C.}~\bibnamefont {Silberhorn}},\ }\href {\doibase
  10.1103/PhysRevLett.104.050502} {\bibfield  {journal} {\bibinfo  {journal}
  {Physical Review Letters}\ }\textbf {\bibinfo {volume} {104}},\ \bibinfo
  {pages} {050502} (\bibinfo {year} {2010})}\BibitemShut {NoStop}%
\bibitem [{\citenamefont {Schreiber}\ \emph {et~al.}(2011)\citenamefont
  {Schreiber}, \citenamefont {Cassemiro}, \citenamefont {Poto{\v{c}}ek},
  \citenamefont {G{\'{a}}bris}, \citenamefont {Jex},\ and\ \citenamefont
  {Silberhorn}}]{Schreiber2011}%
  \BibitemOpen
  \bibfield  {author} {\bibinfo {author} {\bibfnamefont {A.}~\bibnamefont
  {Schreiber}}, \bibinfo {author} {\bibfnamefont {K.~N.}\ \bibnamefont
  {Cassemiro}}, \bibinfo {author} {\bibfnamefont {V.}~\bibnamefont
  {Poto{\v{c}}ek}}, \bibinfo {author} {\bibfnamefont {A.}~\bibnamefont
  {G{\'{a}}bris}}, \bibinfo {author} {\bibfnamefont {I.}~\bibnamefont {Jex}}, \
  and\ \bibinfo {author} {\bibfnamefont {C.}~\bibnamefont {Silberhorn}},\
  }\href {\doibase 10.1103/PhysRevLett.106.180403} {\bibfield  {journal}
  {\bibinfo  {journal} {Physical Review Letters}\ }\textbf {\bibinfo {volume}
  {106}},\ \bibinfo {pages} {180403} (\bibinfo {year} {2011})}\BibitemShut
  {NoStop}%
\bibitem [{\citenamefont {Schreiber}\ \emph {et~al.}(2012)\citenamefont
  {Schreiber}, \citenamefont {Gabris}, \citenamefont {Rohde}, \citenamefont
  {Laiho}, \citenamefont {Stefanak}, \citenamefont {Potocek}, \citenamefont
  {Hamilton}, \citenamefont {Jex},\ and\ \citenamefont
  {Silberhorn}}]{Schreiber2012}%
  \BibitemOpen
  \bibfield  {author} {\bibinfo {author} {\bibfnamefont {A.}~\bibnamefont
  {Schreiber}}, \bibinfo {author} {\bibfnamefont {A.}~\bibnamefont {Gabris}},
  \bibinfo {author} {\bibfnamefont {P.~P.}\ \bibnamefont {Rohde}}, \bibinfo
  {author} {\bibfnamefont {K.}~\bibnamefont {Laiho}}, \bibinfo {author}
  {\bibfnamefont {M.}~\bibnamefont {Stefanak}}, \bibinfo {author}
  {\bibfnamefont {V.}~\bibnamefont {Potocek}}, \bibinfo {author} {\bibfnamefont
  {C.}~\bibnamefont {Hamilton}}, \bibinfo {author} {\bibfnamefont
  {I.}~\bibnamefont {Jex}}, \ and\ \bibinfo {author} {\bibfnamefont
  {C.}~\bibnamefont {Silberhorn}},\ }\href {\doibase 10.1126/science.1218448}
  {\bibfield  {journal} {\bibinfo  {journal} {Science}\ }\textbf {\bibinfo
  {volume} {336}},\ \bibinfo {pages} {55} (\bibinfo {year} {2012})}\BibitemShut
  {NoStop}%
\bibitem [{\citenamefont {Nitsche}\ \emph {et~al.}(2016)\citenamefont
  {Nitsche}, \citenamefont {Elster}, \citenamefont {Novotn{\'{y}}},
  \citenamefont {G{\'{a}}bris}, \citenamefont {Jex}, \citenamefont
  {Barkhofen},\ and\ \citenamefont {Silberhorn}}]{Nitsche2016}%
  \BibitemOpen
  \bibfield  {author} {\bibinfo {author} {\bibfnamefont {T.}~\bibnamefont
  {Nitsche}}, \bibinfo {author} {\bibfnamefont {F.}~\bibnamefont {Elster}},
  \bibinfo {author} {\bibfnamefont {J.}~\bibnamefont {Novotn{\'{y}}}}, \bibinfo
  {author} {\bibfnamefont {A.}~\bibnamefont {G{\'{a}}bris}}, \bibinfo {author}
  {\bibfnamefont {I.}~\bibnamefont {Jex}}, \bibinfo {author} {\bibfnamefont
  {S.}~\bibnamefont {Barkhofen}}, \ and\ \bibinfo {author} {\bibfnamefont
  {C.}~\bibnamefont {Silberhorn}},\ }\href {\doibase
  10.1088/1367-2630/18/6/063017} {\bibfield  {journal} {\bibinfo  {journal}
  {New Journal of Physics}\ }\textbf {\bibinfo {volume} {18}},\ \bibinfo
  {pages} {063017} (\bibinfo {year} {2016})}\BibitemShut {NoStop}%
\bibitem [{\citenamefont {Barkhofen}\ \emph {et~al.}(2017)\citenamefont
  {Barkhofen}, \citenamefont {Nitsche}, \citenamefont {Elster}, \citenamefont
  {Lorz}, \citenamefont {G{\'{a}}bris}, \citenamefont {Jex},\ and\
  \citenamefont {Silberhorn}}]{Barkhofen2017}%
  \BibitemOpen
  \bibfield  {author} {\bibinfo {author} {\bibfnamefont {S.}~\bibnamefont
  {Barkhofen}}, \bibinfo {author} {\bibfnamefont {T.}~\bibnamefont {Nitsche}},
  \bibinfo {author} {\bibfnamefont {F.}~\bibnamefont {Elster}}, \bibinfo
  {author} {\bibfnamefont {L.}~\bibnamefont {Lorz}}, \bibinfo {author}
  {\bibfnamefont {A.}~\bibnamefont {G{\'{a}}bris}}, \bibinfo {author}
  {\bibfnamefont {I.}~\bibnamefont {Jex}}, \ and\ \bibinfo {author}
  {\bibfnamefont {C.}~\bibnamefont {Silberhorn}},\ }\href {\doibase
  10.1103/PhysRevA.96.033846} {\bibfield  {journal} {\bibinfo  {journal}
  {Physical Review A}\ }\textbf {\bibinfo {volume} {96}},\ \bibinfo {pages}
  {033846} (\bibinfo {year} {2017})}\BibitemShut {NoStop}%
\bibitem [{\citenamefont {Chen}\ \emph {et~al.}(2018)\citenamefont {Chen},
  \citenamefont {Ding}, \citenamefont {Qin}, \citenamefont {He}, \citenamefont
  {Luo}, \citenamefont {Chen}, \citenamefont {Liu}, \citenamefont {Wang},
  \citenamefont {Zhang}, \citenamefont {Li}, \citenamefont {You}, \citenamefont
  {Wang}, \citenamefont {Wang}, \citenamefont {Sanders}, \citenamefont {Lu},\
  and\ \citenamefont {Pan}}]{Chen2018}%
  \BibitemOpen
  \bibfield  {author} {\bibinfo {author} {\bibfnamefont {C.}~\bibnamefont
  {Chen}}, \bibinfo {author} {\bibfnamefont {X.}~\bibnamefont {Ding}}, \bibinfo
  {author} {\bibfnamefont {J.}~\bibnamefont {Qin}}, \bibinfo {author}
  {\bibfnamefont {Y.}~\bibnamefont {He}}, \bibinfo {author} {\bibfnamefont
  {Y.-H.}\ \bibnamefont {Luo}}, \bibinfo {author} {\bibfnamefont {M.-C.}\
  \bibnamefont {Chen}}, \bibinfo {author} {\bibfnamefont {C.}~\bibnamefont
  {Liu}}, \bibinfo {author} {\bibfnamefont {X.-L.}\ \bibnamefont {Wang}},
  \bibinfo {author} {\bibfnamefont {W.-J.}\ \bibnamefont {Zhang}}, \bibinfo
  {author} {\bibfnamefont {H.}~\bibnamefont {Li}}, \bibinfo {author}
  {\bibfnamefont {L.-X.}\ \bibnamefont {You}}, \bibinfo {author} {\bibfnamefont
  {Z.}~\bibnamefont {Wang}}, \bibinfo {author} {\bibfnamefont {D.-W.}\
  \bibnamefont {Wang}}, \bibinfo {author} {\bibfnamefont {B.~C.}\ \bibnamefont
  {Sanders}}, \bibinfo {author} {\bibfnamefont {C.-Y.}\ \bibnamefont {Lu}}, \
  and\ \bibinfo {author} {\bibfnamefont {J.-W.}\ \bibnamefont {Pan}},\ }\href
  {\doibase 10.1103/PhysRevLett.121.100502} {\bibfield  {journal} {\bibinfo
  {journal} {Physical Review Letters}\ }\textbf {\bibinfo {volume} {121}},\
  \bibinfo {pages} {100502} (\bibinfo {year} {2018})}\BibitemShut {NoStop}%
\bibitem [{\citenamefont {Hafezi}\ \emph {et~al.}(2011)\citenamefont {Hafezi},
  \citenamefont {Demler}, \citenamefont {Lukin},\ and\ \citenamefont
  {Taylor}}]{Hafezi2011}%
  \BibitemOpen
  \bibfield  {author} {\bibinfo {author} {\bibfnamefont {M.}~\bibnamefont
  {Hafezi}}, \bibinfo {author} {\bibfnamefont {E.~A.}\ \bibnamefont {Demler}},
  \bibinfo {author} {\bibfnamefont {M.~D.}\ \bibnamefont {Lukin}}, \ and\
  \bibinfo {author} {\bibfnamefont {J.~M.}\ \bibnamefont {Taylor}},\ }\href
  {\doibase 10.1038/nphys2063} {\bibfield  {journal} {\bibinfo  {journal}
  {Nature Physics}\ }\textbf {\bibinfo {volume} {7}},\ \bibinfo {pages} {907}
  (\bibinfo {year} {2011})}\BibitemShut {NoStop}%
\bibitem [{\citenamefont {Mittal}\ \emph {et~al.}(2014)\citenamefont {Mittal},
  \citenamefont {Fan}, \citenamefont {Faez}, \citenamefont {Migdall},
  \citenamefont {Taylor},\ and\ \citenamefont {Hafezi}}]{Mittal2014}%
  \BibitemOpen
  \bibfield  {author} {\bibinfo {author} {\bibfnamefont {S.}~\bibnamefont
  {Mittal}}, \bibinfo {author} {\bibfnamefont {J.}~\bibnamefont {Fan}},
  \bibinfo {author} {\bibfnamefont {S.}~\bibnamefont {Faez}}, \bibinfo {author}
  {\bibfnamefont {A.}~\bibnamefont {Migdall}}, \bibinfo {author} {\bibfnamefont
  {J.~M.}\ \bibnamefont {Taylor}}, \ and\ \bibinfo {author} {\bibfnamefont
  {M.}~\bibnamefont {Hafezi}},\ }\href {\doibase
  10.1103/PhysRevLett.113.087403} {\bibfield  {journal} {\bibinfo  {journal}
  {Physical Review Letters}\ }\textbf {\bibinfo {volume} {113}},\ \bibinfo
  {pages} {087403} (\bibinfo {year} {2014})}\BibitemShut {NoStop}%
\bibitem [{\citenamefont {Rechtsman}\ \emph {et~al.}(2013)\citenamefont
  {Rechtsman}, \citenamefont {Zeuner}, \citenamefont {Plotnik}, \citenamefont
  {Lumer}, \citenamefont {Podolsky}, \citenamefont {Dreisow}, \citenamefont
  {Nolte}, \citenamefont {Segev},\ and\ \citenamefont
  {Szameit}}]{Rechtsman2013}%
  \BibitemOpen
  \bibfield  {author} {\bibinfo {author} {\bibfnamefont {M.~C.}\ \bibnamefont
  {Rechtsman}}, \bibinfo {author} {\bibfnamefont {J.~M.}\ \bibnamefont
  {Zeuner}}, \bibinfo {author} {\bibfnamefont {Y.}~\bibnamefont {Plotnik}},
  \bibinfo {author} {\bibfnamefont {Y.}~\bibnamefont {Lumer}}, \bibinfo
  {author} {\bibfnamefont {D.}~\bibnamefont {Podolsky}}, \bibinfo {author}
  {\bibfnamefont {F.}~\bibnamefont {Dreisow}}, \bibinfo {author} {\bibfnamefont
  {S.}~\bibnamefont {Nolte}}, \bibinfo {author} {\bibfnamefont
  {M.}~\bibnamefont {Segev}}, \ and\ \bibinfo {author} {\bibfnamefont
  {A.}~\bibnamefont {Szameit}},\ }\href {\doibase 10.1038/nature12066}
  {\bibfield  {journal} {\bibinfo  {journal} {Nature}\ }\textbf {\bibinfo
  {volume} {496}},\ \bibinfo {pages} {196} (\bibinfo {year}
  {2013})}\BibitemShut {NoStop}%
\bibitem [{\citenamefont {Fang}\ \emph {et~al.}(2012)\citenamefont {Fang},
  \citenamefont {Yu},\ and\ \citenamefont {Fan}}]{Fang2012}%
  \BibitemOpen
  \bibfield  {author} {\bibinfo {author} {\bibfnamefont {K.}~\bibnamefont
  {Fang}}, \bibinfo {author} {\bibfnamefont {Z.}~\bibnamefont {Yu}}, \ and\
  \bibinfo {author} {\bibfnamefont {S.}~\bibnamefont {Fan}},\ }\href {\doibase
  10.1038/nphoton.2012.236} {\bibfield  {journal} {\bibinfo  {journal} {Nature
  Photonics}\ }\textbf {\bibinfo {volume} {6}},\ \bibinfo {pages} {782}
  (\bibinfo {year} {2012})}\BibitemShut {NoStop}%
\bibitem [{\citenamefont {Ozawa}\ \emph {et~al.}(2019)\citenamefont {Ozawa},
  \citenamefont {Price}, \citenamefont {Amo}, \citenamefont {Goldman},
  \citenamefont {Hafezi}, \citenamefont {Lu}, \citenamefont {Rechtsman},
  \citenamefont {Schuster}, \citenamefont {Simon}, \citenamefont {Zilberberg},\
  and\ \citenamefont {Carusotto}}]{Ozawa2019}%
  \BibitemOpen
  \bibfield  {author} {\bibinfo {author} {\bibfnamefont {T.}~\bibnamefont
  {Ozawa}}, \bibinfo {author} {\bibfnamefont {H.~M.}\ \bibnamefont {Price}},
  \bibinfo {author} {\bibfnamefont {A.}~\bibnamefont {Amo}}, \bibinfo {author}
  {\bibfnamefont {N.}~\bibnamefont {Goldman}}, \bibinfo {author} {\bibfnamefont
  {M.}~\bibnamefont {Hafezi}}, \bibinfo {author} {\bibfnamefont
  {L.}~\bibnamefont {Lu}}, \bibinfo {author} {\bibfnamefont {M.~C.}\
  \bibnamefont {Rechtsman}}, \bibinfo {author} {\bibfnamefont {D.}~\bibnamefont
  {Schuster}}, \bibinfo {author} {\bibfnamefont {J.}~\bibnamefont {Simon}},
  \bibinfo {author} {\bibfnamefont {O.}~\bibnamefont {Zilberberg}}, \ and\
  \bibinfo {author} {\bibfnamefont {I.}~\bibnamefont {Carusotto}},\ }\href
  {\doibase 10.1103/RevModPhys.91.015006} {\bibfield  {journal} {\bibinfo
  {journal} {Reviews of Modern Physics}\ }\textbf {\bibinfo {volume} {91}},\
  \bibinfo {pages} {015006} (\bibinfo {year} {2019})}\BibitemShut {NoStop}%
\bibitem [{\citenamefont {Hu}\ \emph {et~al.}(2015)\citenamefont {Hu},
  \citenamefont {Pillay}, \citenamefont {Wu}, \citenamefont {Pasek},
  \citenamefont {Shum},\ and\ \citenamefont {Chong}}]{Hu2015}%
  \BibitemOpen
  \bibfield  {author} {\bibinfo {author} {\bibfnamefont {W.}~\bibnamefont
  {Hu}}, \bibinfo {author} {\bibfnamefont {J.~C.}\ \bibnamefont {Pillay}},
  \bibinfo {author} {\bibfnamefont {K.}~\bibnamefont {Wu}}, \bibinfo {author}
  {\bibfnamefont {M.}~\bibnamefont {Pasek}}, \bibinfo {author} {\bibfnamefont
  {P.~P.}\ \bibnamefont {Shum}}, \ and\ \bibinfo {author} {\bibfnamefont
  {Y.~D.}\ \bibnamefont {Chong}},\ }\href {\doibase 10.1103/PhysRevX.5.011012}
  {\bibfield  {journal} {\bibinfo  {journal} {Physical Review X}\ }\textbf
  {\bibinfo {volume} {5}},\ \bibinfo {pages} {011012} (\bibinfo {year}
  {2015})}\BibitemShut {NoStop}%
\bibitem [{\citenamefont {D'Errico}\ \emph {et~al.}(2018)\citenamefont
  {D'Errico}, \citenamefont {Cardano}, \citenamefont {Maffei}, \citenamefont
  {Dauphin}, \citenamefont {Barboza}, \citenamefont {Esposito}, \citenamefont
  {Piccirillo}, \citenamefont {Lewenstein}, \citenamefont {Massignan},\ and\
  \citenamefont {Marrucci}}]{derrico2018twodimensional}%
  \BibitemOpen
  \bibfield  {author} {\bibinfo {author} {\bibfnamefont {A.}~\bibnamefont
  {D'Errico}}, \bibinfo {author} {\bibfnamefont {F.}~\bibnamefont {Cardano}},
  \bibinfo {author} {\bibfnamefont {M.}~\bibnamefont {Maffei}}, \bibinfo
  {author} {\bibfnamefont {A.}~\bibnamefont {Dauphin}}, \bibinfo {author}
  {\bibfnamefont {R.}~\bibnamefont {Barboza}}, \bibinfo {author} {\bibfnamefont
  {C.}~\bibnamefont {Esposito}}, \bibinfo {author} {\bibfnamefont
  {B.}~\bibnamefont {Piccirillo}}, \bibinfo {author} {\bibfnamefont
  {M.}~\bibnamefont {Lewenstein}}, \bibinfo {author} {\bibfnamefont
  {P.}~\bibnamefont {Massignan}}, \ and\ \bibinfo {author} {\bibfnamefont
  {L.}~\bibnamefont {Marrucci}},\ }\href@noop {} {\enquote {\bibinfo {title}
  {Two-dimensional topological quantum walks in the momentum space of
  structured light},}\ } (\bibinfo {year} {2018}),\ \Eprint
  {http://arxiv.org/abs/1811.04001} {arXiv:1811.04001 [quant-ph]} \BibitemShut
  {NoStop}%
\bibitem [{\citenamefont {Bromberg}\ \emph {et~al.}(2010)\citenamefont
  {Bromberg}, \citenamefont {Lahini},\ and\ \citenamefont
  {Silberberg}}]{Bromberg2010}%
  \BibitemOpen
  \bibfield  {author} {\bibinfo {author} {\bibfnamefont {Y.}~\bibnamefont
  {Bromberg}}, \bibinfo {author} {\bibfnamefont {Y.}~\bibnamefont {Lahini}}, \
  and\ \bibinfo {author} {\bibfnamefont {Y.}~\bibnamefont {Silberberg}},\
  }\href {\doibase 10.1103/PhysRevLett.105.263604} {\bibfield  {journal}
  {\bibinfo  {journal} {Physical Review Letters}\ }\textbf {\bibinfo {volume}
  {105}},\ \bibinfo {pages} {263604} (\bibinfo {year} {2010})}\BibitemShut
  {NoStop}%
\bibitem [{\citenamefont {Corrielli}\ \emph {et~al.}(2013)\citenamefont
  {Corrielli}, \citenamefont {Crespi}, \citenamefont {{Della Valle}},
  \citenamefont {Longhi},\ and\ \citenamefont {Osellame}}]{Corrielli2013}%
  \BibitemOpen
  \bibfield  {author} {\bibinfo {author} {\bibfnamefont {G.}~\bibnamefont
  {Corrielli}}, \bibinfo {author} {\bibfnamefont {A.}~\bibnamefont {Crespi}},
  \bibinfo {author} {\bibfnamefont {G.}~\bibnamefont {{Della Valle}}}, \bibinfo
  {author} {\bibfnamefont {S.}~\bibnamefont {Longhi}}, \ and\ \bibinfo {author}
  {\bibfnamefont {R.}~\bibnamefont {Osellame}},\ }\href {\doibase
  10.1038/ncomms2578} {\bibfield  {journal} {\bibinfo  {journal} {Nature
  Communications}\ }\textbf {\bibinfo {volume} {4}},\ \bibinfo {pages} {1555}
  (\bibinfo {year} {2013})}\BibitemShut {NoStop}%
\bibitem [{\citenamefont {Witthaut}\ \emph {et~al.}(2004)\citenamefont
  {Witthaut}, \citenamefont {Keck}, \citenamefont {Korsch},\ and\ \citenamefont
  {Mossmann}}]{Witthaut2004}%
  \BibitemOpen
  \bibfield  {author} {\bibinfo {author} {\bibfnamefont {D.}~\bibnamefont
  {Witthaut}}, \bibinfo {author} {\bibfnamefont {F.}~\bibnamefont {Keck}},
  \bibinfo {author} {\bibfnamefont {H.~J.}\ \bibnamefont {Korsch}}, \ and\
  \bibinfo {author} {\bibfnamefont {S.}~\bibnamefont {Mossmann}},\ }\href
  {\doibase 10.1088/1367-2630/6/1/041} {\bibfield  {journal} {\bibinfo
  {journal} {New Journal of Physics}\ }\textbf {\bibinfo {volume} {6}},\
  \bibinfo {pages} {41} (\bibinfo {year} {2004})}\BibitemShut {NoStop}%
\bibitem [{\citenamefont {Bloch}(1929)}]{Bloch1929}%
  \BibitemOpen
  \bibfield  {author} {\bibinfo {author} {\bibfnamefont {F.}~\bibnamefont
  {Bloch}},\ }\href {\doibase 10.1007/BF01339455} {\bibfield  {journal}
  {\bibinfo  {journal} {Zeitschrift f�r Physik}\ }\textbf {\bibinfo {volume}
  {52}},\ \bibinfo {pages} {555} (\bibinfo {year} {1929})}\BibitemShut
  {NoStop}%
\bibitem [{\citenamefont {Haller}\ \emph {et~al.}(2010)\citenamefont {Haller},
  \citenamefont {Hart}, \citenamefont {Mark}, \citenamefont {Danzl},
  \citenamefont {Reichs{\"{o}}llner},\ and\ \citenamefont
  {N{\"{a}}gerl}}]{Haller2010}%
  \BibitemOpen
  \bibfield  {author} {\bibinfo {author} {\bibfnamefont {E.}~\bibnamefont
  {Haller}}, \bibinfo {author} {\bibfnamefont {R.}~\bibnamefont {Hart}},
  \bibinfo {author} {\bibfnamefont {M.~J.}\ \bibnamefont {Mark}}, \bibinfo
  {author} {\bibfnamefont {J.~G.}\ \bibnamefont {Danzl}}, \bibinfo {author}
  {\bibfnamefont {L.}~\bibnamefont {Reichs{\"{o}}llner}}, \ and\ \bibinfo
  {author} {\bibfnamefont {H.-C.}\ \bibnamefont {N{\"{a}}gerl}},\ }\href
  {\doibase 10.1103/PhysRevLett.104.200403} {\bibfield  {journal} {\bibinfo
  {journal} {Physical Review Letters}\ }\textbf {\bibinfo {volume} {104}},\
  \bibinfo {pages} {200403} (\bibinfo {year} {2010})}\BibitemShut {NoStop}%
\bibitem [{\citenamefont {{Ben Dahan}}\ \emph {et~al.}(1996)\citenamefont {{Ben
  Dahan}}, \citenamefont {Peik}, \citenamefont {Reichel}, \citenamefont
  {Castin},\ and\ \citenamefont {Salomon}}]{BenDahan1996}%
  \BibitemOpen
  \bibfield  {author} {\bibinfo {author} {\bibfnamefont {M.}~\bibnamefont {{Ben
  Dahan}}}, \bibinfo {author} {\bibfnamefont {E.}~\bibnamefont {Peik}},
  \bibinfo {author} {\bibfnamefont {J.}~\bibnamefont {Reichel}}, \bibinfo
  {author} {\bibfnamefont {Y.}~\bibnamefont {Castin}}, \ and\ \bibinfo {author}
  {\bibfnamefont {C.}~\bibnamefont {Salomon}},\ }\href {\doibase
  10.1103/PhysRevLett.76.4508} {\bibfield  {journal} {\bibinfo  {journal}
  {Physical Review Letters}\ }\textbf {\bibinfo {volume} {76}},\ \bibinfo
  {pages} {4508} (\bibinfo {year} {1996})}\BibitemShut {NoStop}%
\bibitem [{\citenamefont {Pertsch}\ \emph {et~al.}(1999)\citenamefont
  {Pertsch}, \citenamefont {Dannberg}, \citenamefont {Elflein}, \citenamefont
  {Br{\"{a}}uer},\ and\ \citenamefont {Lederer}}]{Pertsch1999}%
  \BibitemOpen
  \bibfield  {author} {\bibinfo {author} {\bibfnamefont {T.}~\bibnamefont
  {Pertsch}}, \bibinfo {author} {\bibfnamefont {P.}~\bibnamefont {Dannberg}},
  \bibinfo {author} {\bibfnamefont {W.}~\bibnamefont {Elflein}}, \bibinfo
  {author} {\bibfnamefont {A.}~\bibnamefont {Br{\"{a}}uer}}, \ and\ \bibinfo
  {author} {\bibfnamefont {F.}~\bibnamefont {Lederer}},\ }\href {\doibase
  10.1103/PhysRevLett.83.4752} {\bibfield  {journal} {\bibinfo  {journal}
  {Physical Review Letters}\ }\textbf {\bibinfo {volume} {83}},\ \bibinfo
  {pages} {4752} (\bibinfo {year} {1999})}\BibitemShut {NoStop}%
\bibitem [{\citenamefont {Morandotti}\ \emph {et~al.}(1999)\citenamefont
  {Morandotti}, \citenamefont {Peschel}, \citenamefont {Aitchison},
  \citenamefont {Eisenberg},\ and\ \citenamefont
  {Silberberg}}]{Morandotti1999}%
  \BibitemOpen
  \bibfield  {author} {\bibinfo {author} {\bibfnamefont {R.}~\bibnamefont
  {Morandotti}}, \bibinfo {author} {\bibfnamefont {U.}~\bibnamefont {Peschel}},
  \bibinfo {author} {\bibfnamefont {J.~S.}\ \bibnamefont {Aitchison}}, \bibinfo
  {author} {\bibfnamefont {H.~S.}\ \bibnamefont {Eisenberg}}, \ and\ \bibinfo
  {author} {\bibfnamefont {Y.}~\bibnamefont {Silberberg}},\ }\href {\doibase
  10.1103/PhysRevLett.83.4756} {\bibfield  {journal} {\bibinfo  {journal}
  {Physical Review Letters}\ }\textbf {\bibinfo {volume} {83}},\ \bibinfo
  {pages} {4756} (\bibinfo {year} {1999})}\BibitemShut {NoStop}%
\bibitem [{\citenamefont {Lenz}\ \emph {et~al.}(1999)\citenamefont {Lenz},
  \citenamefont {Talanina},\ and\ \citenamefont {de~Sterke}}]{Lenz1999}%
  \BibitemOpen
  \bibfield  {author} {\bibinfo {author} {\bibfnamefont {G.}~\bibnamefont
  {Lenz}}, \bibinfo {author} {\bibfnamefont {I.}~\bibnamefont {Talanina}}, \
  and\ \bibinfo {author} {\bibfnamefont {C.~M.}\ \bibnamefont {de~Sterke}},\
  }\href {\doibase 10.1103/PhysRevLett.83.963} {\bibfield  {journal} {\bibinfo
  {journal} {Physical Review Letters}\ }\textbf {\bibinfo {volume} {83}},\
  \bibinfo {pages} {963} (\bibinfo {year} {1999})}\BibitemShut {NoStop}%
\bibitem [{\citenamefont {Trompeter}\ \emph {et~al.}(2006)\citenamefont
  {Trompeter}, \citenamefont {Krolikowski}, \citenamefont {Neshev},
  \citenamefont {Desyatnikov}, \citenamefont {Sukhorukov}, \citenamefont
  {Kivshar}, \citenamefont {Pertsch}, \citenamefont {Peschel},\ and\
  \citenamefont {Lederer}}]{Trompeter2006}%
  \BibitemOpen
  \bibfield  {author} {\bibinfo {author} {\bibfnamefont {H.}~\bibnamefont
  {Trompeter}}, \bibinfo {author} {\bibfnamefont {W.}~\bibnamefont
  {Krolikowski}}, \bibinfo {author} {\bibfnamefont {D.~N.}\ \bibnamefont
  {Neshev}}, \bibinfo {author} {\bibfnamefont {A.~S.}\ \bibnamefont
  {Desyatnikov}}, \bibinfo {author} {\bibfnamefont {A.~A.}\ \bibnamefont
  {Sukhorukov}}, \bibinfo {author} {\bibfnamefont {Y.~S.}\ \bibnamefont
  {Kivshar}}, \bibinfo {author} {\bibfnamefont {T.}~\bibnamefont {Pertsch}},
  \bibinfo {author} {\bibfnamefont {U.}~\bibnamefont {Peschel}}, \ and\
  \bibinfo {author} {\bibfnamefont {F.}~\bibnamefont {Lederer}},\ }\href
  {\doibase 10.1103/PhysRevLett.96.053903} {\bibfield  {journal} {\bibinfo
  {journal} {Physical Review Letters}\ }\textbf {\bibinfo {volume} {96}},\
  \bibinfo {pages} {053903} (\bibinfo {year} {2006})}\BibitemShut {NoStop}%
\bibitem [{\citenamefont {Longhi}(2006)}]{Longhi2006}%
  \BibitemOpen
  \bibfield  {author} {\bibinfo {author} {\bibfnamefont {S.}~\bibnamefont
  {Longhi}},\ }\href {\doibase 10.1209/epl/i2006-10301-8} {\bibfield  {journal}
  {\bibinfo  {journal} {Europhysics Letters (EPL)}\ }\textbf {\bibinfo {volume}
  {76}},\ \bibinfo {pages} {416} (\bibinfo {year} {2006})}\BibitemShut
  {NoStop}%
\bibitem [{\citenamefont {Dreisow}\ \emph {et~al.}(2009)\citenamefont
  {Dreisow}, \citenamefont {Szameit}, \citenamefont {Heinrich}, \citenamefont
  {Pertsch}, \citenamefont {Nolte}, \citenamefont {T{\"{u}}nnermann},\ and\
  \citenamefont {Longhi}}]{Dreisow2009}%
  \BibitemOpen
  \bibfield  {author} {\bibinfo {author} {\bibfnamefont {F.}~\bibnamefont
  {Dreisow}}, \bibinfo {author} {\bibfnamefont {A.}~\bibnamefont {Szameit}},
  \bibinfo {author} {\bibfnamefont {M.}~\bibnamefont {Heinrich}}, \bibinfo
  {author} {\bibfnamefont {T.}~\bibnamefont {Pertsch}}, \bibinfo {author}
  {\bibfnamefont {S.}~\bibnamefont {Nolte}}, \bibinfo {author} {\bibfnamefont
  {A.}~\bibnamefont {T{\"{u}}nnermann}}, \ and\ \bibinfo {author}
  {\bibfnamefont {S.}~\bibnamefont {Longhi}},\ }\href {\doibase
  10.1103/PhysRevLett.102.076802} {\bibfield  {journal} {\bibinfo  {journal}
  {Physical Review Letters}\ }\textbf {\bibinfo {volume} {102}},\ \bibinfo
  {pages} {076802} (\bibinfo {year} {2009})}\BibitemShut {NoStop}%
\bibitem [{\citenamefont {Levy}\ and\ \citenamefont {Kumar}(2010)}]{Levy2010}%
  \BibitemOpen
  \bibfield  {author} {\bibinfo {author} {\bibfnamefont {M.}~\bibnamefont
  {Levy}}\ and\ \bibinfo {author} {\bibfnamefont {P.}~\bibnamefont {Kumar}},\
  }\href {\doibase 10.1364/OL.35.003147} {\bibfield  {journal} {\bibinfo
  {journal} {Optics Letters}\ }\textbf {\bibinfo {volume} {35}},\ \bibinfo
  {pages} {3147} (\bibinfo {year} {2010})}\BibitemShut {NoStop}%
\bibitem [{\citenamefont {Xue}\ \emph {et~al.}(2015)\citenamefont {Xue},
  \citenamefont {Zhang}, \citenamefont {Qin}, \citenamefont {Zhan},
  \citenamefont {Bian}, \citenamefont {Li},\ and\ \citenamefont
  {Sanders}}]{Xue_2015}%
  \BibitemOpen
  \bibfield  {author} {\bibinfo {author} {\bibfnamefont {P.}~\bibnamefont
  {Xue}}, \bibinfo {author} {\bibfnamefont {R.}~\bibnamefont {Zhang}}, \bibinfo
  {author} {\bibfnamefont {H.}~\bibnamefont {Qin}}, \bibinfo {author}
  {\bibfnamefont {X.}~\bibnamefont {Zhan}}, \bibinfo {author} {\bibfnamefont
  {Z.~H.}\ \bibnamefont {Bian}}, \bibinfo {author} {\bibfnamefont
  {J.}~\bibnamefont {Li}}, \ and\ \bibinfo {author} {\bibfnamefont {B.~C.}\
  \bibnamefont {Sanders}},\ }\href {\doibase 10.1103/PhysRevLett.114.140502}
  {\bibfield  {journal} {\bibinfo  {journal} {Physical Review Letters}\
  }\textbf {\bibinfo {volume} {114}},\ \bibinfo {pages} {140502} (\bibinfo
  {year} {2015})}\BibitemShut {NoStop}%
\bibitem [{\citenamefont {Cedzich}\ and\ \citenamefont
  {Werner}(2016)}]{Cedzich2016}%
  \BibitemOpen
  \bibfield  {author} {\bibinfo {author} {\bibfnamefont {C.}~\bibnamefont
  {Cedzich}}\ and\ \bibinfo {author} {\bibfnamefont {R.~F.}\ \bibnamefont
  {Werner}},\ }\href {\doibase 10.1103/PhysRevA.93.032329} {\bibfield
  {journal} {\bibinfo  {journal} {Physical Review A}\ }\textbf {\bibinfo
  {volume} {93}},\ \bibinfo {pages} {032329} (\bibinfo {year}
  {2016})}\BibitemShut {NoStop}%
\bibitem [{\citenamefont {Peschel}\ \emph {et~al.}(2008)\citenamefont
  {Peschel}, \citenamefont {Bersch},\ and\ \citenamefont
  {Onishchukov}}]{Peschel2008}%
  \BibitemOpen
  \bibfield  {author} {\bibinfo {author} {\bibfnamefont {U.}~\bibnamefont
  {Peschel}}, \bibinfo {author} {\bibfnamefont {C.}~\bibnamefont {Bersch}}, \
  and\ \bibinfo {author} {\bibfnamefont {G.}~\bibnamefont {Onishchukov}},\
  }\href {\doibase 10.2478/s11534-008-0095-0} {\bibfield  {journal} {\bibinfo
  {journal} {Open Physics}\ }\textbf {\bibinfo {volume} {6}} (\bibinfo {year}
  {2008}),\ 10.2478/s11534-008-0095-0}\BibitemShut {NoStop}%
\bibitem [{\citenamefont {Bersch}\ \emph {et~al.}(2009)\citenamefont {Bersch},
  \citenamefont {Onishchukov},\ and\ \citenamefont {Peschel}}]{Bersch2009}%
  \BibitemOpen
  \bibfield  {author} {\bibinfo {author} {\bibfnamefont {C.}~\bibnamefont
  {Bersch}}, \bibinfo {author} {\bibfnamefont {G.}~\bibnamefont {Onishchukov}},
  \ and\ \bibinfo {author} {\bibfnamefont {U.}~\bibnamefont {Peschel}},\ }\href
  {\doibase 10.1364/OL.34.002372} {\bibfield  {journal} {\bibinfo  {journal}
  {Optics Letters}\ }\textbf {\bibinfo {volume} {34}},\ \bibinfo {pages} {2372}
  (\bibinfo {year} {2009})}\BibitemShut {NoStop}%
\bibitem [{\citenamefont {Yuan}\ and\ \citenamefont {Fan}(2016)}]{Yuan2016v2}%
  \BibitemOpen
  \bibfield  {author} {\bibinfo {author} {\bibfnamefont {L.}~\bibnamefont
  {Yuan}}\ and\ \bibinfo {author} {\bibfnamefont {S.}~\bibnamefont {Fan}},\
  }\href {\doibase 10.1364/OPTICA.3.001014} {\bibfield  {journal} {\bibinfo
  {journal} {Optica}\ }\textbf {\bibinfo {volume} {3}},\ \bibinfo {pages}
  {1014} (\bibinfo {year} {2016})}\BibitemShut {NoStop}%
\bibitem [{sup()}]{supplem}%
  \BibitemOpen
  \href@noop {} {}\bibinfo {note} {{See Supplemental Material at [URL will be
  inserted by publisher] for methods and theoretical derivations}}\BibitemShut
  {NoStop}%
\bibitem [{\citenamefont {Cedzich}\ \emph {et~al.}(2013)\citenamefont
  {Cedzich}, \citenamefont {Ryb{\'{a}}r}, \citenamefont {Werner}, \citenamefont
  {Alberti}, \citenamefont {Genske},\ and\ \citenamefont
  {Werner}}]{Cedzich2013}%
  \BibitemOpen
  \bibfield  {author} {\bibinfo {author} {\bibfnamefont {C.}~\bibnamefont
  {Cedzich}}, \bibinfo {author} {\bibfnamefont {T.}~\bibnamefont
  {Ryb{\'{a}}r}}, \bibinfo {author} {\bibfnamefont {A.~H.}\ \bibnamefont
  {Werner}}, \bibinfo {author} {\bibfnamefont {A.}~\bibnamefont {Alberti}},
  \bibinfo {author} {\bibfnamefont {M.}~\bibnamefont {Genske}}, \ and\ \bibinfo
  {author} {\bibfnamefont {R.~F.}\ \bibnamefont {Werner}},\ }\href {\doibase
  10.1103/PhysRevLett.111.160601} {\bibfield  {journal} {\bibinfo  {journal}
  {Physical Review Letters}\ }\textbf {\bibinfo {volume} {111}},\ \bibinfo
  {pages} {160601} (\bibinfo {year} {2013})},\ \Eprint
  {http://arxiv.org/abs/1302.2081} {arXiv:1302.2081} \BibitemShut {NoStop}%
\bibitem [{\citenamefont {Genske}\ \emph {et~al.}(2013)\citenamefont {Genske},
  \citenamefont {Alt}, \citenamefont {Steffen}, \citenamefont {Werner},
  \citenamefont {Werner}, \citenamefont {Meschede},\ and\ \citenamefont
  {Alberti}}]{Genske2013}%
  \BibitemOpen
  \bibfield  {author} {\bibinfo {author} {\bibfnamefont {M.}~\bibnamefont
  {Genske}}, \bibinfo {author} {\bibfnamefont {W.}~\bibnamefont {Alt}},
  \bibinfo {author} {\bibfnamefont {A.}~\bibnamefont {Steffen}}, \bibinfo
  {author} {\bibfnamefont {A.~H.}\ \bibnamefont {Werner}}, \bibinfo {author}
  {\bibfnamefont {R.~F.}\ \bibnamefont {Werner}}, \bibinfo {author}
  {\bibfnamefont {D.}~\bibnamefont {Meschede}}, \ and\ \bibinfo {author}
  {\bibfnamefont {A.}~\bibnamefont {Alberti}},\ }\href {\doibase
  10.1103/PhysRevLett.110.190601} {\bibfield  {journal} {\bibinfo  {journal}
  {Physical Review Letters}\ }\textbf {\bibinfo {volume} {110}},\ \bibinfo
  {pages} {190601} (\bibinfo {year} {2013})}\BibitemShut {NoStop}%
\bibitem [{\citenamefont {Bru}\ \emph {et~al.}(2016)\citenamefont {Bru},
  \citenamefont {Hinarejos}, \citenamefont {Silva}, \citenamefont
  {de~Valc{\'{a}}rcel},\ and\ \citenamefont {Rold{\'{a}}n}}]{Bru2016}%
  \BibitemOpen
  \bibfield  {author} {\bibinfo {author} {\bibfnamefont {L.~A.}\ \bibnamefont
  {Bru}}, \bibinfo {author} {\bibfnamefont {M.}~\bibnamefont {Hinarejos}},
  \bibinfo {author} {\bibfnamefont {F.}~\bibnamefont {Silva}}, \bibinfo
  {author} {\bibfnamefont {G.~J.}\ \bibnamefont {de~Valc{\'{a}}rcel}}, \ and\
  \bibinfo {author} {\bibfnamefont {E.}~\bibnamefont {Rold{\'{a}}n}},\ }\href
  {\doibase 10.1103/PhysRevA.93.032333} {\bibfield  {journal} {\bibinfo
  {journal} {Physical Review A}\ }\textbf {\bibinfo {volume} {93}},\ \bibinfo
  {pages} {032333} (\bibinfo {year} {2016})}\BibitemShut {NoStop}%
\bibitem [{\citenamefont {Cedzich}\ \emph {et~al.}(2019)\citenamefont
  {Cedzich}, \citenamefont {Geib}, \citenamefont {Werner},\ and\ \citenamefont
  {Werner}}]{Cedzich2019}%
  \BibitemOpen
  \bibfield  {author} {\bibinfo {author} {\bibfnamefont {C.}~\bibnamefont
  {Cedzich}}, \bibinfo {author} {\bibfnamefont {T.}~\bibnamefont {Geib}},
  \bibinfo {author} {\bibfnamefont {A.~H.}\ \bibnamefont {Werner}}, \ and\
  \bibinfo {author} {\bibfnamefont {R.~F.}\ \bibnamefont {Werner}},\ }\href
  {\doibase 10.1063/1.5054894} {\bibfield  {journal} {\bibinfo  {journal}
  {Journal of Mathematical Physics}\ }\textbf {\bibinfo {volume} {60}},\
  \bibinfo {pages} {012107} (\bibinfo {year} {2019})},\ \Eprint
  {http://arxiv.org/abs/1808.10850} {arXiv:1808.10850} \BibitemShut {NoStop}%
\bibitem [{\citenamefont {Chalabi}\ \emph {et~al.}(2019)\citenamefont
  {Chalabi}, \citenamefont {Barik}, \citenamefont {Mittal}, \citenamefont
  {Murphy}, \citenamefont {Hafezi},\ and\ \citenamefont {Waks}}]{Chalabi2019}%
  \BibitemOpen
  \bibfield  {author} {\bibinfo {author} {\bibfnamefont {H.}~\bibnamefont
  {Chalabi}}, \bibinfo {author} {\bibfnamefont {S.}~\bibnamefont {Barik}},
  \bibinfo {author} {\bibfnamefont {S.}~\bibnamefont {Mittal}}, \bibinfo
  {author} {\bibfnamefont {T.~E.}\ \bibnamefont {Murphy}}, \bibinfo {author}
  {\bibfnamefont {M.}~\bibnamefont {Hafezi}}, \ and\ \bibinfo {author}
  {\bibfnamefont {E.}~\bibnamefont {Waks}},\ }\href {\doibase
  10.1103/PhysRevLett.123.150503} {\bibfield  {journal} {\bibinfo  {journal}
  {Physical Review Letters}\ }\textbf {\bibinfo {volume} {123}},\ \bibinfo
  {pages} {150503} (\bibinfo {year} {2019})}\BibitemShut {NoStop}%
\bibitem [{\citenamefont {Dunlap}\ and\ \citenamefont
  {Kenkre}(1986)}]{Dunlap1986}%
  \BibitemOpen
  \bibfield  {author} {\bibinfo {author} {\bibfnamefont {D.~H.}\ \bibnamefont
  {Dunlap}}\ and\ \bibinfo {author} {\bibfnamefont {V.~M.}\ \bibnamefont
  {Kenkre}},\ }\href {\doibase 10.1103/PhysRevB.34.3625} {\bibfield  {journal}
  {\bibinfo  {journal} {Physical Review B}\ }\textbf {\bibinfo {volume} {34}},\
  \bibinfo {pages} {3625} (\bibinfo {year} {1986})}\BibitemShut {NoStop}%
\bibitem [{\citenamefont {Lenz}\ \emph {et~al.}(2003)\citenamefont {Lenz},
  \citenamefont {Parker}, \citenamefont {Wanke},\ and\ \citenamefont
  {de~Sterke}}]{Lenz2003}%
  \BibitemOpen
  \bibfield  {author} {\bibinfo {author} {\bibfnamefont {G.}~\bibnamefont
  {Lenz}}, \bibinfo {author} {\bibfnamefont {R.}~\bibnamefont {Parker}},
  \bibinfo {author} {\bibfnamefont {M.}~\bibnamefont {Wanke}}, \ and\ \bibinfo
  {author} {\bibfnamefont {C.}~\bibnamefont {de~Sterke}},\ }\href {\doibase
  10.1016/S0030-4018(03)01172-6} {\bibfield  {journal} {\bibinfo  {journal}
  {Optics Communications}\ }\textbf {\bibinfo {volume} {218}},\ \bibinfo
  {pages} {87} (\bibinfo {year} {2003})}\BibitemShut {NoStop}%
\bibitem [{\citenamefont {Longhi}\ \emph {et~al.}(2006)\citenamefont {Longhi},
  \citenamefont {Marangoni}, \citenamefont {Lobino}, \citenamefont {Ramponi},
  \citenamefont {Laporta}, \citenamefont {Cianci},\ and\ \citenamefont
  {Foglietti}}]{Longhi2006v2}%
  \BibitemOpen
  \bibfield  {author} {\bibinfo {author} {\bibfnamefont {S.}~\bibnamefont
  {Longhi}}, \bibinfo {author} {\bibfnamefont {M.}~\bibnamefont {Marangoni}},
  \bibinfo {author} {\bibfnamefont {M.}~\bibnamefont {Lobino}}, \bibinfo
  {author} {\bibfnamefont {R.}~\bibnamefont {Ramponi}}, \bibinfo {author}
  {\bibfnamefont {P.}~\bibnamefont {Laporta}}, \bibinfo {author} {\bibfnamefont
  {E.}~\bibnamefont {Cianci}}, \ and\ \bibinfo {author} {\bibfnamefont
  {V.}~\bibnamefont {Foglietti}},\ }\href {\doibase
  10.1103/PhysRevLett.96.243901} {\bibfield  {journal} {\bibinfo  {journal}
  {Physical Review Letters}\ }\textbf {\bibinfo {volume} {96}},\ \bibinfo
  {pages} {243901} (\bibinfo {year} {2006})}\BibitemShut {NoStop}%
\bibitem [{\citenamefont {Iyer}\ \emph {et~al.}(2007)\citenamefont {Iyer},
  \citenamefont {{Stewart Aitchison}}, \citenamefont {Wan}, \citenamefont
  {Dignam},\ and\ \citenamefont {{Martijn de Sterke}}}]{Iyer2007}%
  \BibitemOpen
  \bibfield  {author} {\bibinfo {author} {\bibfnamefont {R.}~\bibnamefont
  {Iyer}}, \bibinfo {author} {\bibfnamefont {J.}~\bibnamefont {{Stewart
  Aitchison}}}, \bibinfo {author} {\bibfnamefont {J.}~\bibnamefont {Wan}},
  \bibinfo {author} {\bibfnamefont {M.~M.}\ \bibnamefont {Dignam}}, \ and\
  \bibinfo {author} {\bibfnamefont {C.}~\bibnamefont {{Martijn de Sterke}}},\
  }\href {\doibase 10.1364/OE.15.003212} {\bibfield  {journal} {\bibinfo
  {journal} {Optics Express}\ }\textbf {\bibinfo {volume} {15}},\ \bibinfo
  {pages} {3212} (\bibinfo {year} {2007})}\BibitemShut {NoStop}%
\bibitem [{\citenamefont {Joushaghani}\ \emph {et~al.}(2009)\citenamefont
  {Joushaghani}, \citenamefont {Iyer}, \citenamefont {Poon}, \citenamefont
  {Aitchison}, \citenamefont {de~Sterke}, \citenamefont {Wan},\ and\
  \citenamefont {Dignam}}]{Joushaghani2009}%
  \BibitemOpen
  \bibfield  {author} {\bibinfo {author} {\bibfnamefont {A.}~\bibnamefont
  {Joushaghani}}, \bibinfo {author} {\bibfnamefont {R.}~\bibnamefont {Iyer}},
  \bibinfo {author} {\bibfnamefont {J.~K.~S.}\ \bibnamefont {Poon}}, \bibinfo
  {author} {\bibfnamefont {J.~S.}\ \bibnamefont {Aitchison}}, \bibinfo {author}
  {\bibfnamefont {C.~M.}\ \bibnamefont {de~Sterke}}, \bibinfo {author}
  {\bibfnamefont {J.}~\bibnamefont {Wan}}, \ and\ \bibinfo {author}
  {\bibfnamefont {M.~M.}\ \bibnamefont {Dignam}},\ }\href {\doibase
  10.1103/PhysRevLett.103.143903} {\bibfield  {journal} {\bibinfo  {journal}
  {Physical Review Letters}\ }\textbf {\bibinfo {volume} {103}},\ \bibinfo
  {pages} {143903} (\bibinfo {year} {2009})}\BibitemShut {NoStop}%
\bibitem [{\citenamefont {Joushaghani}\ \emph {et~al.}(2012)\citenamefont
  {Joushaghani}, \citenamefont {Iyer}, \citenamefont {Poon}, \citenamefont
  {Aitchison}, \citenamefont {de~Sterke}, \citenamefont {Wan},\ and\
  \citenamefont {Dignam}}]{Joushaghani2012}%
  \BibitemOpen
  \bibfield  {author} {\bibinfo {author} {\bibfnamefont {A.}~\bibnamefont
  {Joushaghani}}, \bibinfo {author} {\bibfnamefont {R.}~\bibnamefont {Iyer}},
  \bibinfo {author} {\bibfnamefont {J.~K.~S.}\ \bibnamefont {Poon}}, \bibinfo
  {author} {\bibfnamefont {J.~S.}\ \bibnamefont {Aitchison}}, \bibinfo {author}
  {\bibfnamefont {C.~M.}\ \bibnamefont {de~Sterke}}, \bibinfo {author}
  {\bibfnamefont {J.}~\bibnamefont {Wan}}, \ and\ \bibinfo {author}
  {\bibfnamefont {M.~M.}\ \bibnamefont {Dignam}},\ }\href {\doibase
  10.1103/PhysRevLett.109.103901} {\bibfield  {journal} {\bibinfo  {journal}
  {Physical Review Letters}\ }\textbf {\bibinfo {volume} {109}},\ \bibinfo
  {pages} {103901} (\bibinfo {year} {2012})}\BibitemShut {NoStop}%
\bibitem [{\citenamefont {Yuan}\ and\ \citenamefont {Fan}(2015)}]{Yuan2015}%
  \BibitemOpen
  \bibfield  {author} {\bibinfo {author} {\bibfnamefont {L.}~\bibnamefont
  {Yuan}}\ and\ \bibinfo {author} {\bibfnamefont {S.}~\bibnamefont {Fan}},\
  }\href {\doibase 10.1103/PhysRevLett.114.243901} {\bibfield  {journal}
  {\bibinfo  {journal} {Physical Review Letters}\ }\textbf {\bibinfo {volume}
  {114}},\ \bibinfo {pages} {243901} (\bibinfo {year} {2015})}\BibitemShut
  {NoStop}%
\bibitem [{\citenamefont {Orre}\ \emph {et~al.}(2019)\citenamefont {Orre},
  \citenamefont {Goldschmidt}, \citenamefont {Deshpande}, \citenamefont
  {Gorshkov}, \citenamefont {Tamma}, \citenamefont {Hafezi},\ and\
  \citenamefont {Mittal}}]{Orre2019}%
  \BibitemOpen
  \bibfield  {author} {\bibinfo {author} {\bibfnamefont {V.~V.}\ \bibnamefont
  {Orre}}, \bibinfo {author} {\bibfnamefont {E.~A.}\ \bibnamefont
  {Goldschmidt}}, \bibinfo {author} {\bibfnamefont {A.}~\bibnamefont
  {Deshpande}}, \bibinfo {author} {\bibfnamefont {A.~V.}\ \bibnamefont
  {Gorshkov}}, \bibinfo {author} {\bibfnamefont {V.}~\bibnamefont {Tamma}},
  \bibinfo {author} {\bibfnamefont {M.}~\bibnamefont {Hafezi}}, \ and\ \bibinfo
  {author} {\bibfnamefont {S.}~\bibnamefont {Mittal}},\ }\href {\doibase
  10.1103/PhysRevLett.123.123603} {\bibfield  {journal} {\bibinfo  {journal}
  {Physical Review Letters}\ }\textbf {\bibinfo {volume} {123}},\ \bibinfo
  {pages} {123603} (\bibinfo {year} {2019})}\BibitemShut {NoStop}%
\bibitem [{\citenamefont {Pichler}\ \emph {et~al.}(2017)\citenamefont
  {Pichler}, \citenamefont {Choi}, \citenamefont {Zoller},\ and\ \citenamefont
  {Lukin}}]{Pichler2017}%
  \BibitemOpen
  \bibfield  {author} {\bibinfo {author} {\bibfnamefont {H.}~\bibnamefont
  {Pichler}}, \bibinfo {author} {\bibfnamefont {S.}~\bibnamefont {Choi}},
  \bibinfo {author} {\bibfnamefont {P.}~\bibnamefont {Zoller}}, \ and\ \bibinfo
  {author} {\bibfnamefont {M.~D.}\ \bibnamefont {Lukin}},\ }\href {\doibase
  10.1073/pnas.1711003114} {\bibfield  {journal} {\bibinfo  {journal}
  {Proceedings of the National Academy of Sciences}\ }\textbf {\bibinfo
  {volume} {114}},\ \bibinfo {pages} {11362} (\bibinfo {year}
  {2017})}\BibitemShut {NoStop}%
\bibitem [{\citenamefont {Chalabi}\ and\ \citenamefont
  {Waks}(2018)}]{Chalabi2018}%
  \BibitemOpen
  \bibfield  {author} {\bibinfo {author} {\bibfnamefont {H.}~\bibnamefont
  {Chalabi}}\ and\ \bibinfo {author} {\bibfnamefont {E.}~\bibnamefont {Waks}},\
  }\href {\doibase 10.1103/PhysRevA.98.063832} {\bibfield  {journal} {\bibinfo
  {journal} {Physical Review A}\ }\textbf {\bibinfo {volume} {98}},\ \bibinfo
  {pages} {063832} (\bibinfo {year} {2018})}\BibitemShut {NoStop}%
\end{thebibliography}%

\onecolumngrid



\section*{Supplementary material}

# Supplemental Material: Guiding and confining of light in a two-dimensional synthetic space using electric fields

Hamidreza Chalabi,[1,2] Sabyasachi Barik,[1,2,3] Sunil Mittal,[1,2]
Thomas E. Murphy,[1,3] Mohammad Hafezi,[1,2,3] and Edo Waks[1,2,3]

[1] Department of Electrical and Computer Engineering and Institute for Research in Electronics and Applied Physics,
University of Maryland, College Park, Maryland 20742, USA
[2] Joint Quantum Institute, University of Maryland, College Park, Maryland 20742, USA
[3] Department of Physics, University of Maryland, College Park, Maryland 20742, USA

We explain the experimental details of this study in the first section of this supplemental material. In this section, we also present the measurement results as well as theoretical predictions for the variation of the quadratic means and norm ones of the quantum walk distributions as functions of the time step. In the second section, we review the theoretical analysis of one-dimensional quantum walks and explain the possibility of observing Bloch oscillations using time-dependent gauge fields in such quantum walks. In the next section, we provide the corresponding theoretical analysis for the case of two-dimensional quantum walks.

## I. EXPERIMENTAL DETAILS:

Figure S1 shows the schematic of the experimental setup, which is composed of similar elements as used in our previous work [S1]. However, the phase modulation patterns used in this study are different from those used earlier. We used a laser diode made by Bookham technology (LC25W5172BA-J34) operating at the C-band of the telecom wavelength (1550 nm) for pulse generation. By modulating this laser, pulses with a power of $10\,mW$ and pulse duration of $2\,ns$ were generated, which were sent into the setup with a repetition rate of $10\,\mu s$. These pulses initiate the quantum walk in our synthetic two-dimensional space. The initial pulse after passing through the first 50/50 beam splitter continues its path through either the 1 m or 2 m fiber length. This leads to different delays that can be conceptually regarded as "choosing" the left or right direction in the $x$ movement by the quantum walker. When reaching the next beam splitter, the pulse chooses either the 130 m or 118 m fiber spool. This determines whether the pulse has gone in the up or down direction in the $y$ movement. The $x$ and $y$ coordinate of the pulses are encoded based on the delay in reaching the detectors. We used 90/10 beam splitters to out-couple 10% of the light for detection purposes. In addition, two phase modulators made by Cybel, LLC (MPZ-LN-10-P-P-FA-FA) were used to apply opposite phases to the right and left moving pulses (i.e., $\pm x$ direction).

We used semiconductor optical amplifiers made by Thorlabs (SOA1117S) to compensate for the loss in the setup

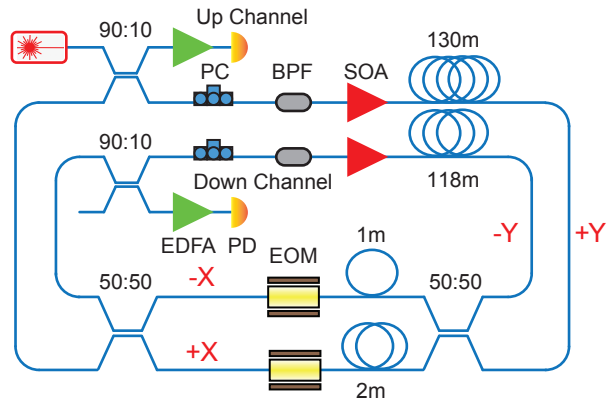

Figure S1. Schematic of the experimental setup for the two-dimensional quantum random walk. PD: photodetector, BPF: band-pass filter, SOA: semiconductor optical amplifier, EDFA: Erbium-doped fiber amplifier, EOM: Electro-optic modulator and PC: polarization controller.

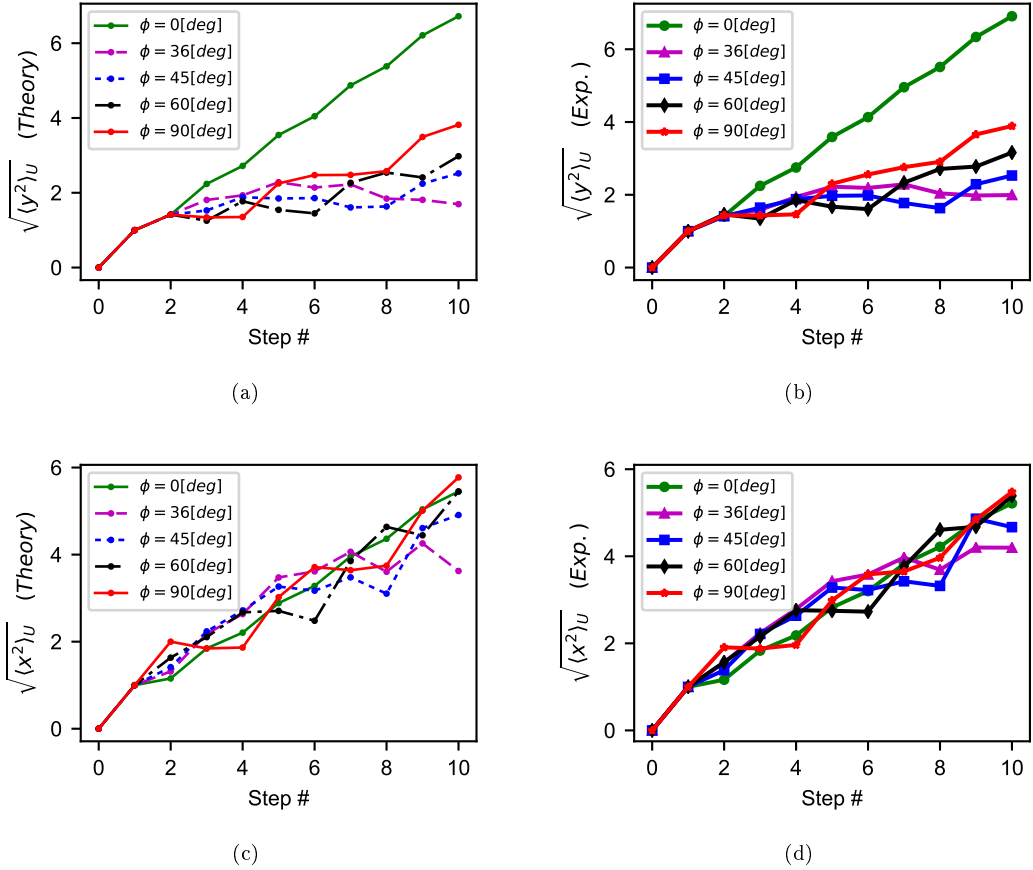

Figure S2. Theoretical values for the quadratic means of (a) $y$ and (c) $x$ as a function of the time step for different gauge field strengths. Experimental values for the quadratic means of (b) $y$ and (d) $x$ as a function of the time step for different gauge field strengths.

by amplifying the pulses without ruining their phase coherence. These amplifiers are only turned on for a fraction of a repetition period at each cycle to avoid over amplifying the background noises. Moreover, in order to filter out background noises, we used narrow band-pass filters ($< 0.3nm$) directly after the semiconductor optical amplifiers. However, the ratio of the optical pulse power at different positions of the synthetic space relative to the noise degrades with increasing time steps due to the diffusion and the added noise. We therefore limited our measurement of the quantum walk distribution to 10 steps because of the observed degradation of the signal-to-noise ratio and the finite ratio of the time delays corresponding to the $x$ and $y$ movements. The distribution of the quantum walk at every time step was determined by measuring the power of the pulses present in each step. We also used polarization controllers in our setup in order to compensate for the polarization changes along the optical fibers.

We measured the quantum walk distribution at different time steps under the application of the proposed time-varying gauge fields with different phases ($\phi = 0\,[deg]$, $\phi = 90\,[deg]$, $\phi = 60\,[deg]$, $\phi = 45\,[deg]$, and $\phi = 36\,[deg]$).

Using the measured probability distributions, we calculated the quadratic means and norm ones of the $x$ and $y$ of the quantum walk distribution. We plotted the variation of these quantities by the time step in Figs. S2 and S3. These quantities reach their local minima after around $2\pi/\phi$ time steps depending on the value of $\phi$ used. This behavior is due to the state revival and the trapping of the quantum walk distribution close to the origin after around $2\pi/\phi$ time steps.

We can also calculate the theoretically predicted values for the quadratic means and norm ones as functions of the time step. As Figs. S2 and S3 show, the measured results are in good agreement with the theoretical predictions.

Based on the measured results, we can also calculate the quantum walk distribution probability at the origin for different applied phases. Using these probabilities, we plotted the revival probability in the main text as a function of the required steps to reach the revival. Indeed, smaller phases lead to higher probability of the revival, as has been

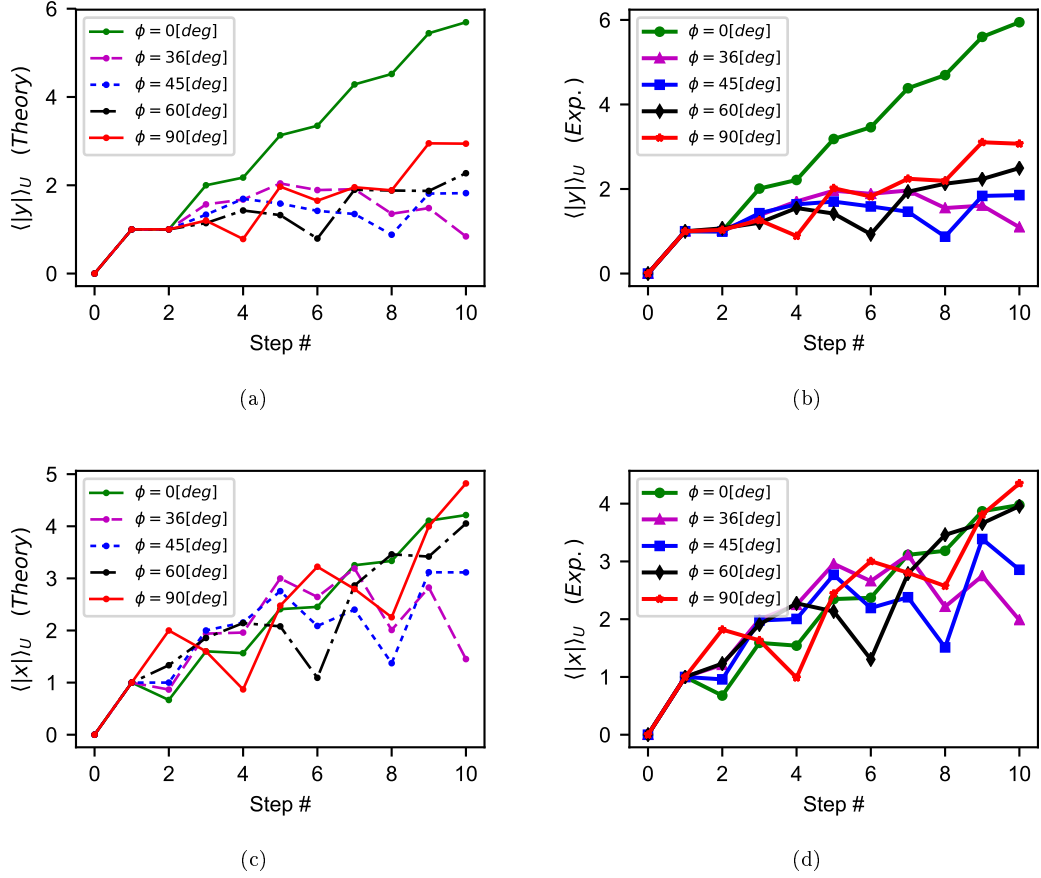

Figure S3. Theoretical values for the norm ones of (a) $y$ and (c) $x$ as a function of the time step for different gauge field strengths. Experimental values for the norm ones of (b) $y$ and (d) $x$ as a function of the time step for different gauge field strengths.

discussed in the main text.

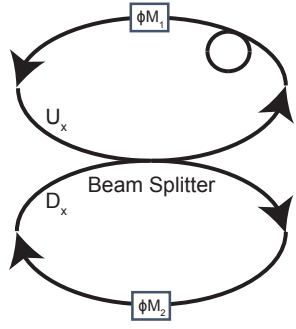

Figure S4. Simplified version of the schematic of a one-dimensional quantum random walk setup

## II. THEORETICAL ANALYSIS FOR ONE-DIMENSIONAL QUANTUM RANDOM WALKS:

In this section, we consider the evolution of pulses (representing the quantum walkers) in one-dimensional quantum random walks. For the theoretical analysis of the quantum walk, we can assume that all the losses in the setup are fully compensated by the amplifiers. Therefore, we can consider an ideal setup as shown in Fig. S4.

Two pulses just before the first beam splitter, $\begin{bmatrix} U_x^{(n)} \\ D_x^{(n)} \end{bmatrix}$, will produce the output pulses as (movement in the $x$ direction):

$$\begin{bmatrix} U_{x+1}^{(n+1)} \\ D_{x-1}^{(n+1)} \end{bmatrix} = \frac{1}{\sqrt{2}} \begin{bmatrix} e^{-i\phi_{n,x}} & 0 \\ 0 & e^{i\phi_{n,x}} \end{bmatrix} \begin{bmatrix} 1 & -1 \\ 1 & 1 \end{bmatrix} \begin{bmatrix} U_x^{(n)} \\ D_x^{(n)} \end{bmatrix}.$$

Therefore, $\begin{bmatrix} U_x^{(n)} \\ D_x^{(n)} \end{bmatrix}$ will produce the following pulses after traversing the beam splitter:

$$U_{x+1}^{(n+1)} = \frac{e^{-i\phi_{n,x}}}{\sqrt{2}} \left( U_x^{(n)} - D_x^{(n)} \right)$$

$$D_{x-1}^{(n+1)} = \frac{e^{i\phi_{n,x}}}{\sqrt{2}} \left( U_x^{(n)} + D_x^{(n)} \right)$$

Alternatively, based on the above results $\begin{bmatrix} U_x^{(n+1)} \\ D_x^{(n+1)} \end{bmatrix}$ can be produced from other pulses as:

$$U_x^{(n+1)} = \frac{e^{-i\phi_{n,x-1}}}{\sqrt{2}} \left( U_{x-1}^{(n)} - D_{x-1}^{(n)} \right)$$

$$D_x^{(n+1)} = \frac{e^{i\phi_{n,x+1}}}{\sqrt{2}} \left( U_{x+1}^{(n)} + D_{x+1}^{(n)} \right)$$

By defining $S_x^{(n)} = U_x^{(n)} + D_x^{(n)}$ and $P_x^{(n)} = U_x^{(n)} - D_x^{(n)}$, the obtained equations can be written as:

$$S_x^{(n+1)} = \frac{e^{i\phi_{n,x+1}}}{\sqrt{2}} S_{x+1}^{(n)} + \frac{e^{-i\phi_{n,x-1}}}{\sqrt{2}} P_{x-1}^{(n)}$$

$$P_x^{(n+1)} = -\frac{e^{i\phi_{n,x+1}}}{\sqrt{2}} S_{x+1}^{(n)} + \frac{e^{-i\phi_{n,x-1}}}{\sqrt{2}} P_{x-1}^{(n)}$$

$$\begin{bmatrix} S_x^{(n+1)} \\ P_x^{(n+1)} \end{bmatrix} = \frac{1}{\sqrt{2}} \begin{bmatrix} e^{i\phi_{n,x+1}} & e^{-i\phi_{n,x-1}} \\ -e^{i\phi_{n,x+1}} & e^{-i\phi_{n,x-1}} \end{bmatrix} \begin{bmatrix} S_{x+1}^{(n)} \\ P_{x-1}^{(n)} \end{bmatrix}$$

By defining $s_{k_x}^{(n)}$ and $p_{k_x}^{(n)}$ as Fourier transforms of $S_x^{(n)}$ and $P_x^{(n)}$, we have:

$$\begin{bmatrix} S_x^{(n)} \\ P_x^{(n)} \end{bmatrix} = \frac{1}{2\pi} \begin{bmatrix} \int_{k_x} s_{k_x}^{(n)} e^{ik_x x} dk_x \\ \int_{k_x} p_{k_x}^{(n)} e^{ik_x x} dk_x \end{bmatrix}$$

Therefore:

$$\begin{bmatrix} \int_{k_x} s_{k_x}^{(n+1)} e^{ik_x x} dk_x \\ \int_{k_x} p_{k_x}^{(n+1)} e^{ik_x x} dk_x \end{bmatrix} = \frac{1}{\sqrt{2}} \begin{bmatrix} e^{i\phi_{n,x+1}} & e^{-i\phi_{n,x-1}} \\ -e^{i\phi_{n,x+1}} & e^{-i\phi_{n,x-1}} \end{bmatrix}$$
$$\times \begin{bmatrix} \int_{k_x} e^{ik_x} s_{k_x}^{(n)} e^{ik_x x} dk_x \\ \int_{k_x} e^{-ik_x} p_{k_x}^{(n)} e^{ik_x x} dk_x \end{bmatrix}$$

This equation can be used to solve the Fourier transforms as functions of the time step. In the following subsections, we solve this equation for two cases of no phase modulation as well as time dependent linear phase modulation.

## A. Zero phase modulation:

For the case of no applied phase, we have:

$$\begin{bmatrix} s_{k_x}^{(n+1)} \\ p_{k_x}^{(n+1)} \end{bmatrix} = \frac{1}{\sqrt{2}} \begin{bmatrix} e^{ik_x} & e^{-ik_x} \\ -e^{ik_x} & e^{-ik_x} \end{bmatrix} \begin{bmatrix} s_{k_x}^{(n)} \\ p_{k_x}^{(n)} \end{bmatrix}$$

Note that the evolution matrix has the determinant of 1. Based on this matrix, the effective Hamiltonian, $H_{eff} = i \log(U)$, is given by:

$$H = \frac{1}{\sqrt{2}} \begin{pmatrix} \sin(k_x) & -ie^{-ik_x} \\ ie^{ik_x} & -\sin(k_x) \end{pmatrix} \frac{\arccos\left(\cos(k_x)/\sqrt{2}\right)}{\sin\left(\arccos\left(\cos(k_x)/\sqrt{2}\right)\right)}$$

The obtained Hamiltonian is hermitian and its eigenvalues are given by:

$$E_\pm = \pm \arccos\left(\cos(k_x)/\sqrt{2}\right)$$

The evolution after $n$ steps is given by:

$$\begin{bmatrix} s_{k_x}^{(n)} \\ p_{k_x}^{(n)} \end{bmatrix} = \frac{1}{\sqrt{2^n}} \begin{bmatrix} e^{ik_x} & e^{-ik_x} \\ -e^{ik_x} & e^{-ik_x} \end{bmatrix}^n \begin{bmatrix} s_{k_x}^{(0)} \\ p_{k_x}^{(0)} \end{bmatrix}$$

Assuming that the whole evolution is caused by a single pulse at the origin ($U_x^{(0)} = \delta(x)$, $D_x^{(0)} = 0$ which is equivalent to $S_x^{(0)} = \delta(x)$, $P_x^{(0)} = \delta(x)$), we have:

$$\begin{bmatrix} s_{k_x}^{(0)} \\ p_{k_x}^{(0)} \end{bmatrix} = \begin{bmatrix} 1 \\ 1 \end{bmatrix}$$

Therefore, for the down channel pulses, $d_{k_x}^{(n)} = 0.5 \left( s_{k_x}^{(n)} - p_{k_x}^{(n)} \right)$, we have:

$$d_{k_x}^{(n)} = \frac{e^{ik_x}}{\sqrt{2}} \frac{\sin \left( n \arccos \left( \cos \left( k_x \right) / \sqrt{2} \right) \right)}{\sin \left( \arccos \left( \cos \left( k_x \right) / \sqrt{2} \right) \right)}$$

All the moments, such as the distribution variances and averages ($\langle x^2 \rangle_D$ and $\langle x \rangle_D$ ) as functions of the time step $n$, can be obtained from the above expression.

Based on the fact that:

$$d_{k_x}^{(n)} = \sum_x D_x^{(n)} e^{-ik_x x}$$

Then:

$$d_{l_x}^{(n)*} d_{k_x}^{(n)} = \sum_p \sum_x e^{-i(k_x x - l_x p)} D_p^{(n)*} D_x^{(n)}$$

Therefore, the following equations can be obtained straightforwardly:

$$P_D = \sum_x \left| D_x^{(n)} \right|^2 = \frac{1}{2\pi} \int_{k_x=0}^{2\pi} \left| d_{k_x}^{(n)} \right|^2 dk_x$$
$$= \frac{1}{4\pi} \int_{k_x=0}^{2\pi} \frac{\sin^2 \left( n \arccos \left( \cos \left( k_x \right) / \sqrt{2} \right) \right)}{\sin^2 \left( \arccos \left( \cos \left( k_x \right) / \sqrt{2} \right) \right)} dk_x$$

$$\langle x \rangle_D = \sum_x x \left| D_x^{(n)} \right|^2 = \frac{1}{2\pi} \int_{k_x=0}^{2\pi} d_{k_x}^{(n)*} \left( i \frac{d}{dk_x} \right) d_{k_x}^{(n)} dk_x = 0$$

$$\langle x^2 \rangle_D = \sum_x x^2 \left| D_x^{(n)} \right|^2 = \frac{1}{2\pi} \int_{k_x=0}^{2\pi} d_{k_x}^{(n)*} \left( i \frac{d}{dk_x} \right)^2 d_{k_x}^{(n)} dk_x$$

In the limit of large $n$, we have:

$$P_D \rightarrow \frac{1}{8\pi} \int_{k_x=0}^{2\pi} \frac{1}{1 - \cos^2 \left( k_x \right) / 2} dk_x = \frac{1}{2\sqrt{2}}$$

$$\langle x \rangle_D \rightarrow 0$$

$$\langle x^2 \rangle_D \rightarrow \frac{n^2}{16\pi} \int_{k_x=0}^{2\pi} \frac{\sin^2 \left( k_x \right)}{\left( 1 - \cos^2 \left( k_x \right) / 2 \right)^2} dk_x = \frac{n^2}{8\sqrt{2}}$$

These results prove that the spatial quadratic mean of the quantum walk distribution varies linearly with the time step. Note that these average values are normalized relative to the total power in the up and down channels. However, they can also be normalized relative to the total power present in the corresponding channel. Since the probabilities of $P_D$ and $P_U$ tend toward constant values, by the latter normalization the asymptotic behavior of the quadratic mean remains linear relative to the time step.

Similarly, we can investigate the up channel:

$$u_{k_x}^{(n)} = 0.5 \left( s_{k_x}^{(n)} + p_{k_x}^{(n)} \right)$$

$$= \cos \left( n \arccos \left( \cos \left( k_x \right) / \sqrt{2} \right) \right) - i \frac{\sin \left( k_x \right)}{\sqrt{2}} \frac{\sin \left( n \arccos \left( \cos \left( k_x \right) / \sqrt{2} \right) \right)}{\sin \left( \arccos \left( \cos \left( k_x \right) / \sqrt{2} \right) \right)}$$

$$P_U = \sum_x \left| U_x^{(n)} \right|^2 = \frac{1}{2\pi} \int_{k_x=0}^{2\pi} \left| u_{k_x}^{(n)} \right|^2 dk_x$$

$$= \frac{1}{2\pi} \int_{k_x=0}^{2\pi} \cos^2 \left( n \arccos \left( \cos \left( k_x \right) / \sqrt{2} \right) \right) dk_x$$

$$+ \frac{1}{4\pi} \int_{k_x=0}^{2\pi} \sin^2 \left( k_x \right) \frac{\sin^2 \left( n \arccos \left( \cos \left( k_x \right) / \sqrt{2} \right) \right)}{\sin^2 \left( \arccos \left( \cos \left( k_x \right) / \sqrt{2} \right) \right)} dk_x$$

$$\langle x \rangle_U = \sum_x x \left| U_x^{(n)} \right|^2 = \frac{1}{2\pi} \int_{k_x=0}^{2\pi} u_{k_x}^{(n)*} \left( i \frac{d}{dk_x} \right) u_{k_x}^{(n)} dk_x$$

$$\langle x^2 \rangle_U = \sum_x x^2 \left| U_x^{(n)} \right|^2 = \frac{1}{2\pi} \int_{k_x=0}^{2\pi} u_{k_x}^{(n)*} \left( i \frac{d}{dk_x} \right)^2 u_{k_x}^{(n)} dk_x$$

In the limit of large $n$, we have:

$$P_U \rightarrow \frac{1}{2} + \frac{1}{8\pi} \int_{k_x=0}^{2\pi} \frac{\sin^2 \left( k_x \right)}{1 - \cos^2 \left( k_x \right) / 2} dk_x = 1 - \frac{1}{2\sqrt{2}}$$

$$\langle x \rangle_U \rightarrow \left( 1 - \frac{\sqrt{2}}{2} \right) n$$

$$\langle x^2 \rangle_U \rightarrow \frac{n^2}{8\pi} \int_{k_x=0}^{2\pi} \frac{\sin^2 \left( k_x \right) \left( \sin^2 \left( k_x \right) + \frac{1}{2} \right)}{\left( 1 - \frac{\cos^2 \left( k_x \right)}{2} \right)^2} dk_x = \left( 1 - \frac{9}{8\sqrt{2}} \right) n^2$$

**B. Time-dependent phase modulation:**

For the case of a time-dependent but coordinate-independent phase modulation, we have:

$$\begin{bmatrix} s_{k_x}^{(n+1)} \\ p_{k_x}^{(n+1)} \end{bmatrix} = \frac{1}{\sqrt{2}} \begin{bmatrix} e^{i(k_x+\phi_n)} & e^{-i(k_x+\phi_n)} \\ -e^{i(k_x+\phi_n)} & e^{-i(k_x+\phi_n)} \end{bmatrix} \begin{bmatrix} s_{k_x}^{(n)} \\ p_{k_x}^{(n)} \end{bmatrix}$$

We consider phase modulations that are linearly varying with the time step, such that $\phi_n = n\phi$. Therefore, by defining $w_{k_x}^{(n)} = s_{k_x-n\phi}^{(n)}$ and $v_{k_x}^{(n)} = p_{k_x-n\phi}^{(n)}$:

$$\begin{bmatrix} w_{k_x+\phi}^{(n+1)} \\ v_{k_x+\phi}^{(n+1)} \end{bmatrix} = \frac{1}{\sqrt{2}} \begin{bmatrix} e^{ik_x} & e^{-ik_x} \\ -e^{ik_x} & e^{-ik_x} \end{bmatrix} \begin{bmatrix} w_{k_x}^{(n)} \\ v_{k_x}^{(n)} \end{bmatrix}$$

The evolution matrix can be obtained for rational values of phase modulations, $\phi/2\pi = p/q$. We then can obtain the pseudo-energy band diagrams from the evolution matrix. In the following, we obtain the eigen-energies for any rational value of phase modulation, straightforwardly.

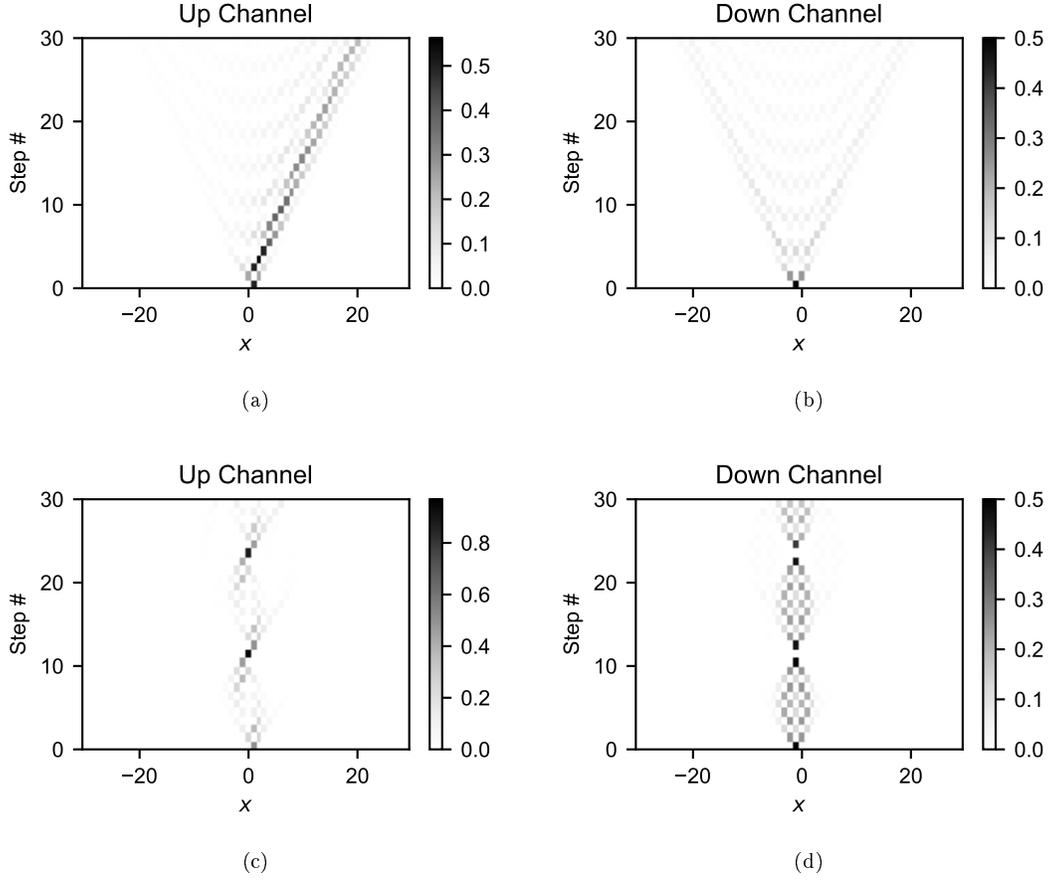

Figure S5. The quantum walk probability distribution in the (a) up and (b) down channels at different time steps under no applied gauge field. The quantum walk probability distribution in the (c) up and (d) down channels at different time steps under a time-varying gauge field with a phase of $\phi = \pi/6$. Under the presence of a nonzero phase, the walker returns back to the origin after a fixed number of steps, which is determined by the applied phase.

By defining $r_{k_x}^{(n)} = \begin{bmatrix} w_{k_x}^{(n)} & v_{k_x}^{(n)} \end{bmatrix}^T$ and

$$
\begin{aligned}
M_{k_x} &= \frac{1}{\sqrt{2}} \begin{bmatrix} e^{ik_x} & e^{-ik_x} \\ -e^{ik_x} & e^{-ik_x} \end{bmatrix} \\
&= \frac{1}{\sqrt{2}} \begin{bmatrix} 1 & 1 \\ -1 & 1 \end{bmatrix} \begin{bmatrix} e^{ik_x} & 0 \\ 0 & e^{-ik_x} \end{bmatrix},
\end{aligned}
$$

we have:

$$
r_{k_x+\phi}^{(n+1)} = M_{k_x} r_{k_x}^{(n)}
$$

After $q$ steps, we have:

$$
r_{k_x}^{(n+q)} = M_{k_x+(q-1)\phi} \dots M_{k_x} r_{k_x}^{(n)}
$$

Therefore, by defining $Y_{k_x}^{(i)} = M_{k_x+(i-1)\phi} \dots M_{k_x}$, the following holds:

$$
M_{k_x} = \frac{1}{\sqrt{2}} \begin{bmatrix} e^{ik_x} & e^{-ik_x} \\ -e^{ik_x} & e^{-ik_x} \end{bmatrix}
$$

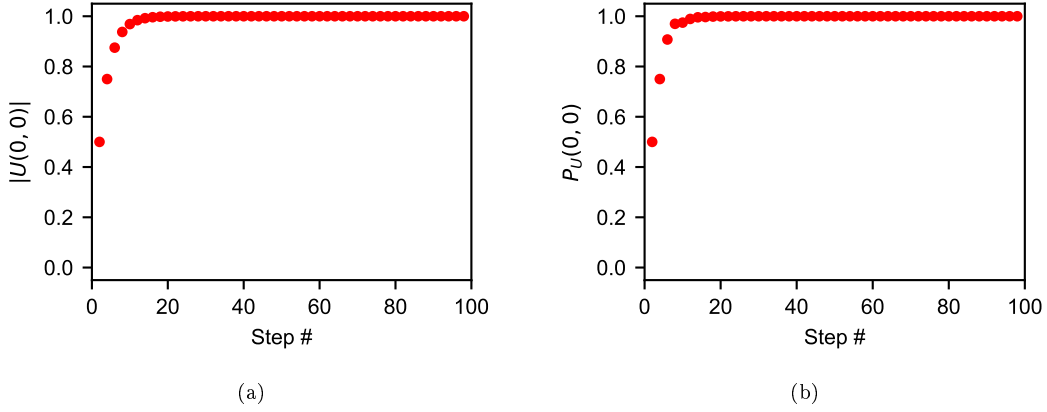

(a)                                                    (b)

Figure S6. (a) The probability amplitude of the quantum walker to be at the origin in the up channel after $q$ steps as a function of $q$. (b) The normalized probability of the quantum walker to return to the origin in the up channel after $q$ steps as a function of $q$.

$$r_{k_x}^{(n+q)} = Y_{k_x}^{(q)} r_{k_x}^{(n)}$$

Since $M$ and $Y$ are unitary matrices, the following equations hold for the eigen-energies:

$$Y_{k_x}^{(q)} r_{k_x}^{(n)} = e^{iqE} r_{k_x}^{(n)}$$

$$Y_{k_x}^{(q)\dagger} r_{k_x}^{(n)} = e^{-iqE} r_{k_x}^{(n)}$$

Consequently, we have:

$$\frac{1}{2} \left( Y_{k_x}^{(q)} + Y_{k_x}^{(q)\dagger} \right) r_{k_x}^{(n)} = \cos\left(qE\right) r_{k_x}^{(n)}$$

It can be verified that not only for $\phi = 2\pi/q$ but also for any $\phi = 2\pi p/q$, with $q$ and $p$ being relatively prime, the following holds:

$$\frac{1}{2} \left( Y_{k_x}^{(q)} + Y_{k_x}^{(q)\dagger} \right) = \left( \left( \cos\left(\frac{\pi q}{2}\right) - \cos\left(qk_x\right) \right) \frac{(-1)^q}{\sqrt{2^q}} - \cos\left(\frac{\pi q}{2}\right) \right) I$$

Therefore, the eigen-energies are given by

$$E_{n,\pm,k_x} = \frac{2n\pi}{q} \pm \frac{1}{q} \arccos\left( \left( \cos\left(\frac{\pi q}{2}\right) - \cos\left(qk_x\right) \right) \frac{(-1)^q}{\sqrt{2^q}} - \cos\left(\frac{\pi q}{2}\right) \right),$$

which explicitly expresses the eigen-energies for any of $2q$ bands.

Based on this analytical expression for the allowed energies, it can be verified that the bandgap does not exist for any rational value of phase modulation.

Since we are interested in the evolution of the quantum walk after $q$ steps, we obtain the following for any $k_x = \frac{\pi}{2} + \frac{2\pi m}{q}$ with $m \in \mathbb{Z}$:

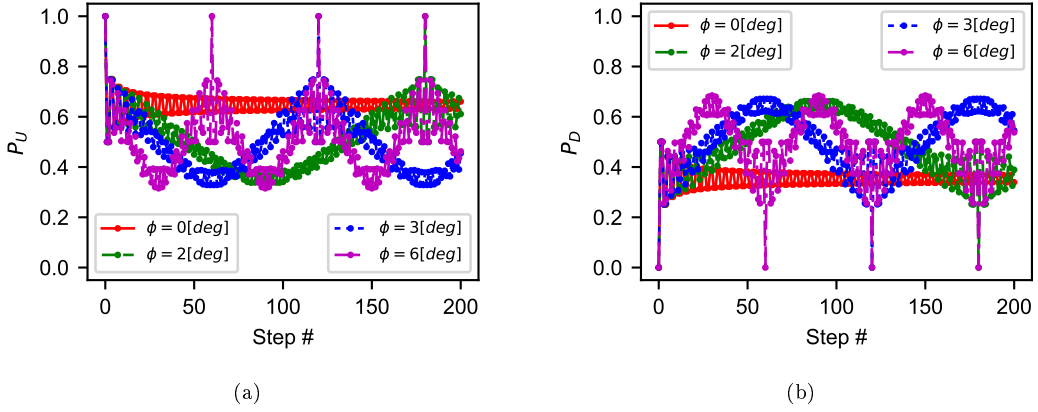

(a)                                                  (b)

Figure S7. (a) The total probability of the quantum walker to be in the up channel and (b) down channel as a function of the time step for different values of phase modulations.

$$qE_{n,\pm,k_x} = \pm \arccos\left[\left(\cos\left(\frac{\pi q}{2}\right) - \cos\left(qk_x\right)\right)\frac{(-1)^q}{\sqrt{2^q}} - \cos\left(\frac{\pi q}{2}\right)\right]$$

$$= \begin{cases} \pi: & q = 4g \\ \pm\pi/2: & q = 4g+1 \\ 0: & q = 4g+2 \\ \pm\pi/2: & q = 4g+3 \end{cases},$$

which corresponds to:

$$e^{-iqE_{n,\pm,k_x}} = \begin{cases} -1: & q = 4g \\ \mp i: & q = 4g+1 \\ 1: & q = 4g+2 \\ \mp i: & q = 4g+3 \end{cases}.$$

We can express any amplitude $\psi$ representing the up channel $U(x,n)$ or the down channel $D(x,n)$ in terms of the initial conditions as:

$$\psi(k_x,n) = \sum_{j=1}^{2q} e^{-iE_{j,k_x}n} A_{j,k_x}$$

Therefore, based on the above expressions, such an amplitude is expressed after $q$ steps for even values of $q$ via:

$$\psi(k_x,q) = s\sum_{j=1}^{2q} e^{ik_x x} A_{j,k_x} = s\psi(k_x,0),$$

in which

$$s = \begin{cases} -1: & q = 4g \\ +1: & q = 4g+2 \end{cases}.$$

This equality holds for any $k_x = \frac{\pi}{2} + \frac{2\pi m}{q}$ with $m \in \mathbb{Z}$ and for other values of $k_x$ it holds by approximation. However, with the increase of $q$, the approximation becomes more and more accurate. Therefore, the following approximation holds, and it becomes exact in the limit of $q \to \infty$:

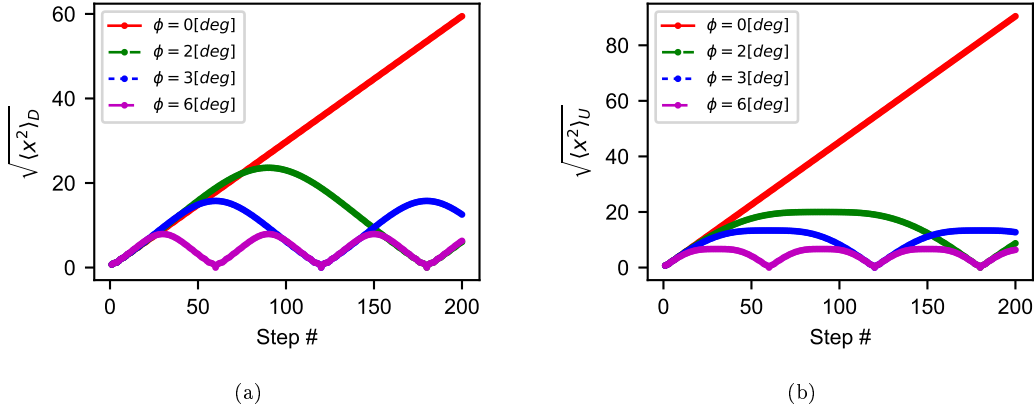

(a)

(b)

Figure S8. The quadratic mean of $x$ for the (a) down channel and (b) up channel as a function of the time step for different values of phase modulations.

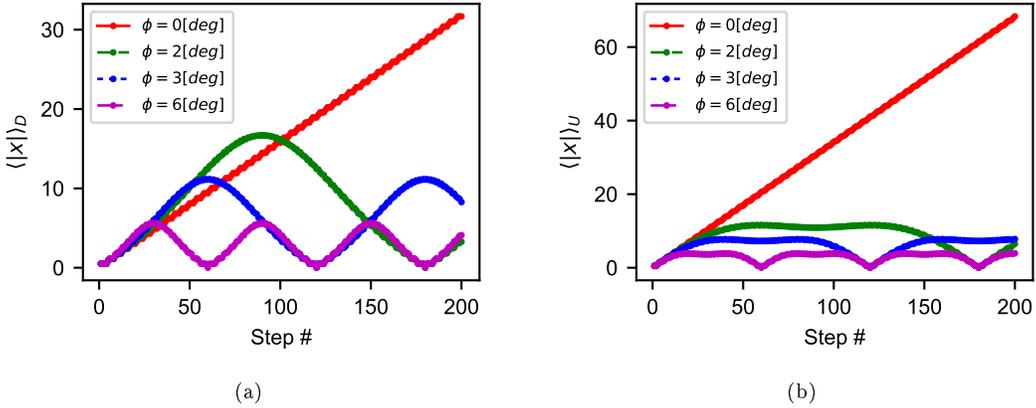

(a)

(b)

Figure S9. The norm one of $x$ for the (a) down channel and (b) up channel as a function of the time step for different values of phase modulations.

$$\psi(x, q) \cong s\psi(x, 0).$$

This proves that revival happens after $q$ steps ($|\psi(x, q)| \cong |\psi(x, 0)|$) for even values of $q$ and it becomes more accurate with the increase of $q$. The variations of the quantum walk probability distribution as a function of the time step under no applied gauge field as well as a linearly time varying gauge field with $\phi = \pi/6$ are shown in Fig. S5. This figure clearly shows the revival caused by Bloch oscillations under the time varying gauge field.

Since we start with a unity pulse at the origin in the up channel, we can plot the probability amplitude of the quantum walker to be at the origin after $q$ steps. We have plotted this amplitude in Fig. S6a. This figure confirms the fact that with the increase of $q$, the quantum walker indeed becomes more localized at the origin and the corresponding probability tends toward unity in the limit of $q \to \infty$. The corresponding normalized probability of the quantum walker to return to the origin after $q$ steps is also shown in Fig. S6b.

We can also consider the total probability of the quantum walker to be in the up channel or down channel as a function of the time step. We have plotted these probabilities in Fig. S7 for different values of phase modulations. As we expect, the probability of the quantum walker to be in the up channel is close to one after $2\pi/\phi$ time steps. It is interesting to note that the minimum value of the probability of the quantum walker to be in the up channel is not zero and instead it is close to $1/2\sqrt{2}$. The probability in the up channel reaches this value after around $\pi/\phi$ time steps.

Figures S8a and S8b summarize the numerical results for the variation of $\sqrt{\langle x^2 \rangle_D}$ and $\sqrt{\langle x^2 \rangle_U}$ with the time step for different values of $\phi$. The obtained results show that for $1 \ll n \ll \pi/\phi$, $\sqrt{\langle x^2 \rangle_D}$ varies as $n/\sqrt{8\sqrt{2}}$ as one would expect from the zero-phase modulation case. Moreover, the numerical results show that for small values of $\phi$, $\sqrt{\langle x^2 \rangle_D}$ reaches the maximum value at $n \cong \pi/\phi$. These results also show that $\langle x^2 \rangle_D$ tends toward zero at $n \cong 2\pi/\phi$, which is consistent with the above analysis. In addition to the quadratic means, we can also look at the variations in the norm ones of the quantum walk distributions. Figures S9a and S9b summarize the numerical results for the variation of $\langle |x| \rangle_D$ and $\langle |x| \rangle_U$ with the time step for different values of $\phi$.

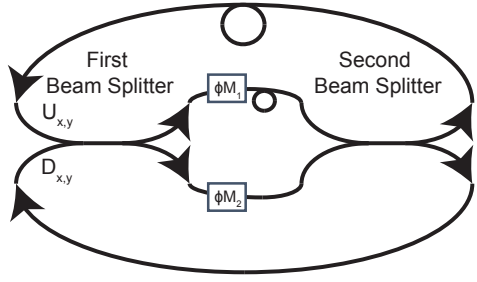

Figure S10. Simplified version of the schematic of a two-dimensional quantum random walk setup

## III. THEORETICAL ANALYSIS FOR TWO-DIMENSIONAL QUANTUM RANDOM WALKS:

In this section, we generalize the concepts that are presented in the previous section to two-dimensional quantum random walks. By assuming that all the losses in the setup are fully compensated by the amplifiers and all the polarization changes are compensated by polarization controllers, the setup can be represented as an ideal setup, as shown in Fig. S10. We have demonstrated in our earlier work how pulses will propagate under arbitrary phase modulation patterns [S1]. By representing the pulses that exist at time step $n$ by $U_{x,y}^{(n)}$ and $D_{x,y}^{(n)}$ in the up and down channels, respectively, we have shown the evolution of the pulses obey the following equations:

$$U_{x,y}^{(n+1)} = \frac{e^{i\phi_{n,y-1}}}{2}\left(U_{x+1,y-1}^{(n)} - D_{x+1,y-1}^{(n)}\right) - \frac{e^{-i\phi_{n,y-1}}}{2}\left(U_{x-1,y-1}^{(n)} + D_{x-1,y-1}^{(n)}\right)$$

$$D_{x,y}^{(n+1)} = \frac{e^{i\phi_{n,y+1}}}{2}\left(U_{x+1,y+1}^{(n)} - D_{x+1,y+1}^{(n)}\right) + \frac{e^{-i\phi_{n,y+1}}}{2}\left(U_{x-1,y+1}^{(n)} + D_{x-1,y+1}^{(n)}\right)$$

By defining $S_{x,y}^{(n)} = U_{x,y}^{(n)} + D_{x,y}^{(n)}$ and $P_{x,y}^{(n)} = U_{x,y}^{(n)} - D_{x,y}^{(n)}$, the obtained equations can be written as:

$$\begin{bmatrix} S_{x,y}^{(n+1)} \\ P_{x,y}^{(n+1)} \end{bmatrix} = \frac{1}{2}\begin{bmatrix} e^{-i\phi_{n,y+1}} & e^{i\phi_{n,y+1}} & -e^{-i\phi_{n,y-1}} & e^{i\phi_{n,y-1}} \\ -e^{-i\phi_{n,y+1}} & -e^{i\phi_{n,y+1}} & -e^{-i\phi_{n,y-1}} & e^{i\phi_{n,y-1}} \end{bmatrix}\begin{bmatrix} S_{x-1,y+1}^{(n)} \\ P_{x+1,y+1}^{(n)} \\ S_{x-1,y-1}^{(n)} \\ P_{x+1,y-1}^{(n)} \end{bmatrix}$$

By defining $s_{k_x,k_y}^{(n)}$ and $p_{k_x,k_y}^{(n)}$ as Fourier transforms of $S_{x,y}^{(n)}$ and $P_{x,y}^{(n)}$, we have:

$$\begin{bmatrix} S_{x,y}^{(n)} \\ P_{x,y}^{(n)} \end{bmatrix} = \frac{1}{4\pi^2}\begin{bmatrix} \int\int_{k_x,k_y} s_{k_x,k_y}^{(n)} e^{ik_x x + ik_y y} dk_x dk_y \\ \int\int_{k_x,k_y} p_{k_x,k_y}^{(n)} e^{ik_x x + ik_y y} dk_x dk_y \end{bmatrix}$$

Therefore:

$$\begin{bmatrix} \int\int_{k_x,k_y} s_{k_x,k_y}^{(n+1)} e^{ik_x x + ik_y y} dk_x dk_y \\ \int\int_{k_x,k_y} p_{k_x,k_y}^{(n+1)} e^{ik_x x + ik_y y} dk_x dk_y \end{bmatrix} = \frac{1}{2}\begin{bmatrix} e^{-i\phi_{n,y+1}} & e^{i\phi_{n,y+1}} & -e^{-i\phi_{n,y-1}} & e^{i\phi_{n,y-1}} \\ -e^{-i\phi_{n,y+1}} & -e^{i\phi_{n,y+1}} & -e^{-i\phi_{n,y-1}} & e^{i\phi_{n,y-1}} \end{bmatrix}$$

$$\times \begin{bmatrix} \int\int_{k_x,k_y} e^{ik_y - ik_x} s_{k_x,k_y}^{(n)} e^{ik_x x + ik_y y} dk_x dk_y \\ \int\int_{k_x,k_y} e^{ik_x + ik_y} p_{k_x,k_y}^{(n)} e^{ik_x x + ik_y y} dk_x dk_y \\ \int\int_{k_x,k_y} e^{-ik_x - ik_y} s_{k_x,k_y}^{(n)} e^{ik_x x + ik_y y} dk_x dk_y \\ \int\int_{k_x,k_y} e^{ik_x - ik_y} p_{k_x,k_y}^{(n)} e^{ik_x x + ik_y y} dk_x dk_y \end{bmatrix}$$

This equation can be used to solve the Fourier transforms as functions of the time step. The evolution for the case of no phase modulation has already been investigated [S1]. Here we focus on the case in which we apply a phase modulation that varies linearly with the time step.

## A. Time-dependent phase modulation:

For the case of time-dependent but coordinate-independent phase modulation, we have:

$$
\begin{bmatrix} s_{k_x,k_y}^{(n+1)} \\ p_{k_x,k_y}^{(n+1)} \end{bmatrix} = \begin{bmatrix} ie^{-i(k_x+\phi_n)}\sin(k_y) & e^{i(k_x+\phi_n)}\cos(k_y) \\ -e^{-i(k_x+\phi_n)}\cos(k_y) & -ie^{i(k_x+\phi_n)}\sin(k_y) \end{bmatrix} \begin{bmatrix} s_{k_x,k_y}^{(n)} \\ p_{k_x,k_y}^{(n)} \end{bmatrix}
$$

Therefore, by defining $w_{k_x,k_y}^{(n)} = s_{k_x-n\phi,k_y}^{(n)}$ and $v_{k_x,k_y}^{(n)} = p_{k_x-n\phi,k_y}^{(n)}$:

$$
\begin{bmatrix} w_{k_x+\phi,k_y}^{(n+1)} \\ v_{k_x+\phi,k_y}^{(n+1)} \end{bmatrix} = \begin{bmatrix} ie^{-ik_x}\sin(k_y) & e^{ik_x}\cos(k_y) \\ -e^{-ik_x}\cos(k_y) & -ie^{ik_x}\sin(k_y) \end{bmatrix} \begin{bmatrix} w_{k_x,k_y}^{(n)} \\ v_{k_x,k_y}^{(n)} \end{bmatrix}
$$

Using this equation, we can obtain a propagation matrix for rational values of phase modulations, $\phi/2\pi = p/q$. In the following, we obtain the eigen-energies for any rational value of phase modulation, straightforwardly.

By defining $r_{k_x,k_y}^{(n)} = \begin{bmatrix} w_{k_x,k_y}^{(n)} & v_{k_x,k_y}^{(n)} \end{bmatrix}^T$ and

$$
\begin{aligned}
M_{k_x,k_y} &= \begin{bmatrix} ie^{-ik_x}\sin(k_y) & e^{ik_x}\cos(k_y) \\ -e^{-ik_x}\cos(k_y) & -ie^{ik_x}\sin(k_y) \end{bmatrix} \\
&= \begin{bmatrix} i\sin(k_y) & \cos(k_y) \\ -\cos(k_y) & -i\sin(k_y) \end{bmatrix} \begin{bmatrix} e^{-ik_x} & 0 \\ 0 & e^{ik_x} \end{bmatrix},
\end{aligned}
$$

we have:

$$
r_{k_x+\phi,k_y}^{(n+1)} = M_{k_x,k_y} r_{k_x,k_y}^{(n)}
$$

After $q$ steps, we have:

$$
r_{k_x,k_y}^{(n+q)} = M_{k_x+(q-1)\phi,k_y} \dots M_{k_x,k_y} r_{k_x,k_y}^{(n)}
$$

Therefore, by defining $Y_{k_x,k_y}^{(i)} = M_{k_x+(i-1)\phi,k_y} \dots M_{k_x,k_y}$, the following holds:

$$
r_{k_x,k_y}^{(n+q)} = Y_{k_x,k_y}^{(q)} r_{k_x,k_y}^{(n)}
$$

Since $M$ and $Y$ are unitary matrices, the following equations hold for the eigen-energies:

$$
Y_{k_x,k_y}^{(q)} r_{k_x,k_y}^{(n)} = e^{iqE} r_{k_x,k_y}^{(n)}
$$

$$
Y_{k_x,k_y}^{(q)\dagger} r_{k_x,k_y}^{(n)} = e^{-iqE} r_{k_x,k_y}^{(n)}
$$

Consequently, we have:

$$
\frac{1}{2}\left(Y_{k_x,k_y}^{(q)} + Y_{k_x,k_y}^{(q)\dagger}\right) r_{k_x,k_y}^{(n)} = \cos(qE) r_{k_x,k_y}^{(n)}
$$

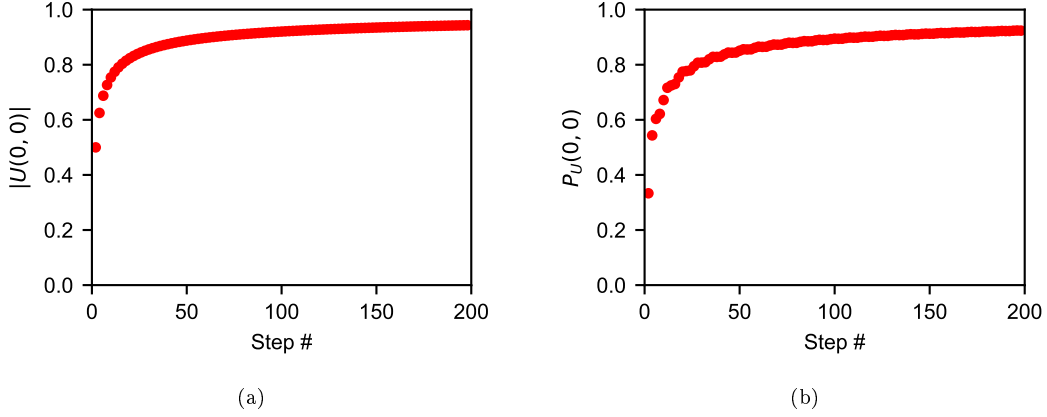

(a)

(b)

Figure S11. (a) The probability amplitude of the quantum walker to be at the origin in the up channel after $q$ steps as a function of $q$. (b) The normalized probability of the quantum walker to return to the origin in the up channel after $q$ steps as a function of $q$.

It can be verified that not only for $\phi = 2\pi/q$ but also for any $\phi = 2\pi p/q$, with $q$ and $p$ being relatively prime, the following holds:

$$\frac{1}{2}\left(Y^{(q)}_{k_x,k_y} + Y^{(q)\dagger}_{k_x,k_y}\right) = \left(\left(\cos\left(\frac{\pi q}{2}\right) - \cos\left(qk_x + \frac{\pi q}{2}\right)\right)\sin^q\left(k_y\right) - \cos\left(\frac{\pi q}{2}\right)\right) I$$

Therefore, the eigen-energies are given by

$$E_{n,\pm,k_x,k_y} = \frac{2n\pi}{q} \pm \frac{1}{q}\arccos\left[\left(\cos\left(\frac{\pi q}{2}\right) - \cos\left(qk_x + \frac{\pi q}{2}\right)\right)\sin^q\left(k_y\right) - \cos\left(\frac{\pi q}{2}\right)\right],$$

which explicitly expresses the eigen-energies for any of $2q$ bands.

Based on this analytical expression for the allowed energies, it can be verified that the bandgap does not exist for any rational value of phase modulation. The energy band diagrams for $\phi = \pi/2$, $\phi = \pi/3$, and $\phi = \pi/4$ have been plotted in Fig. 2 of the main text.

Since we are interested in the evolution of the quantum walk after $q$ steps, we obtain the following for any $k_x = \frac{2\pi m}{q}$ with $m \in \mathbb{Z}$:

$$qE_{n,\pm,k_x,k_y} = \pm\arccos\left[\left(\cos\left(\frac{\pi q}{2}\right) - \cos\left(qk_x + \frac{\pi q}{2}\right)\right)\sin^q\left(k_y\right) - \cos\left(\frac{\pi q}{2}\right)\right]$$

$$= \begin{cases} \pi: & q = 4g \\ \pm\pi/2: & q = 4g+1 \\ 0: & q = 4g+2 \\ \pm\pi/2: & q = 4g+3 \end{cases},$$

which corresponds to:

$$e^{-iqE_{n,\pm,k_x,k_y}} = \begin{cases} -1: & q = 4g \\ \mp i: & q = 4g+1 \\ 1: & q = 4g+2 \\ \mp i: & q = 4g+3 \end{cases}.$$

We can express any amplitude $\psi$ representing the up channel $U(x, y, n)$ or the down channel $D(x, y, n)$ in terms of the initial conditions as:

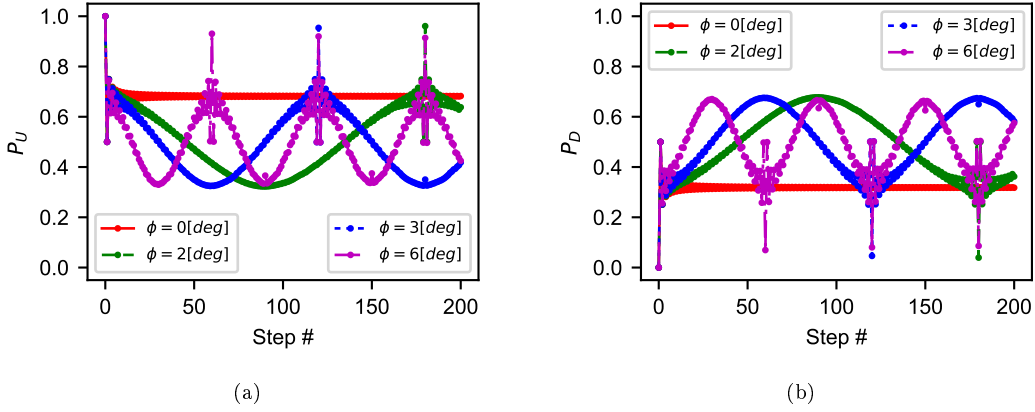

(a)             (b)

Figure S12. (a) The total probability of the quantum walker to be in the up channel and (b) the total probability of the quantum walker to be in the down channel as a function of the time step for different values of phase modulations.

$$\psi\left(k_x, y, n\right) = \sum_{j=1}^{2q} \int_{k_y} e^{-iE_{j,k_x,k_y}n} e^{ik_y y} A_{j,k_x,k_y} dk_y$$

Therefore, based on the above expressions, such an amplitude is expressed after $q$ steps for even values of $q$ via:

$$\psi\left(k_x, y, q\right) = s \sum_{j=1}^{2q} \int_{k_y} e^{ik_y y} A_{j,k_x,k_y} dk_y = s\psi\left(k_x, y, 0\right),$$

in which

$$s = \begin{cases} -1: & q = 4g \\ +1: & q = 4g + 2 \end{cases}.$$

This equality holds for any $k_x = \frac{2\pi m}{q}$ with $m \in \mathbb{Z}$ and for other values of $k_x$ it holds by approximation. However, with the increase of $q$, the approximation becomes more and more accurate. Therefore, the following approximation holds, and it becomes exact in the limit of $q \to \infty$:

$$\psi\left(x, y, q\right) \cong s\psi\left(x, y, 0\right).$$

This proves that revival (caused by Bloch oscillations) happens after $q$ steps ($|\psi\left(x, y, q\right)| \cong |\psi\left(x, y, 0\right)|$) for even values of $q$ and it becomes more accurate with the increase of $q$.

Since we start with a unity pulse at the origin in the up channel, we can plot the probability amplitude of the quantum walker to be at the origin after $q$ steps. We have plotted this amplitude in Fig. S11a. This figure confirms the fact that with the increase of $q$, the quantum walker indeed becomes more trapped at the origin and the corresponding probability tends toward unity in the limit of $q \to \infty$. The corresponding normalized probability of the quantum walker to return to the origin after $q$ steps is also shown in Fig. S11b.

We can also consider the total probability of the quantum walker to be in the up channel or down channel as a function of the time step. We have plotted these probabilities in Fig. S12 for different values of phase modulations. As we expect, the probability of the quantum walker to be in the up channel is close to one after $2\pi/\phi$ time steps. The minimum value of the probability of the quantum walker to be in the up channel is not zero and instead it is close to $1/\pi$. The probability in the up channel reaches this value after around $\pi/\phi$ time steps.

We can also analyze the transient variation of the quantum walk. For this purpose, the obtained propagation equation for $w_{k_x,k_y}^{(n)}$ and $v_{k_x,k_y}^{(n)}$ can be transformed to the following equation:

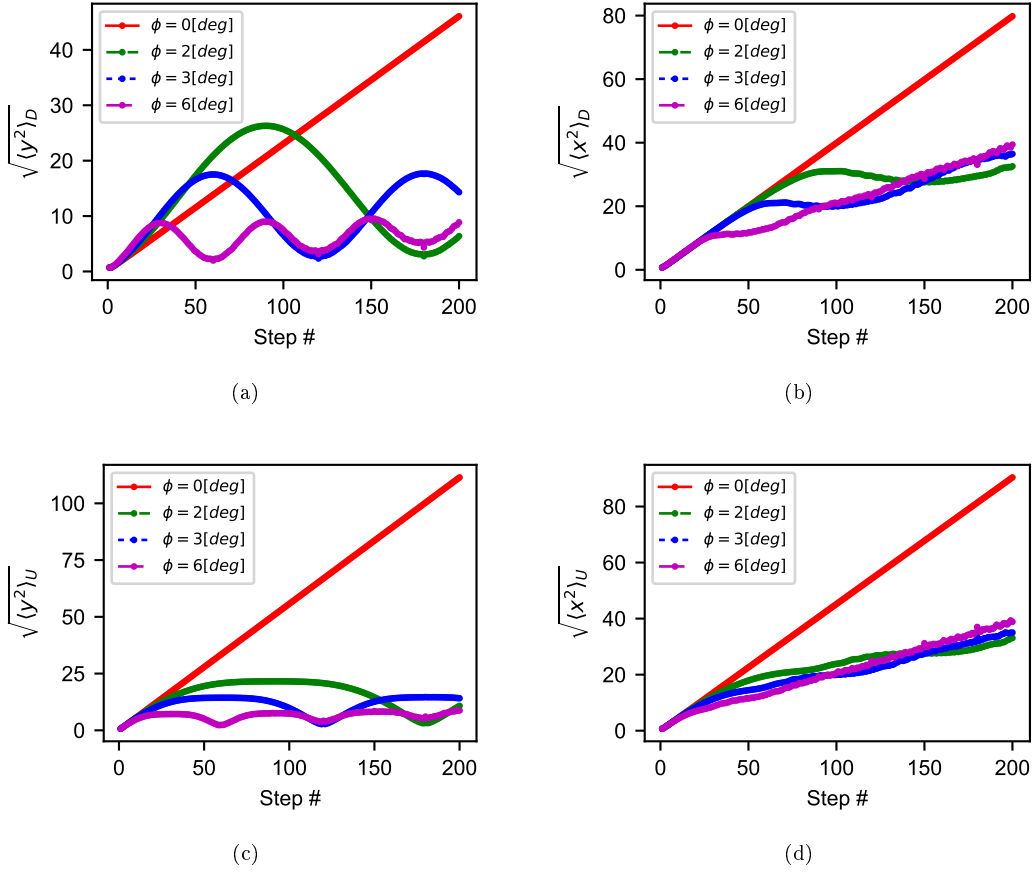

(a)

(b)

(c)

(d)

Figure S13. The quadratic mean of (a) $y$ and (b) $x$ for the down channel as a function of the time step for different phase modulation values. The quadratic mean of (c) $y$ and (d) $x$ for the up channel as a function of the time step for different phase modulation values.

$$
\begin{bmatrix} w_{k_x,k_y}^{(n+1)} + v_{k_x,k_y}^{(n+1)} \\ w_{k_x,k_y}^{(n+1)} - v_{k_x,k_y}^{(n+1)} \end{bmatrix} = \begin{bmatrix} ie^{-ik_y}\sin\left(k_x - \phi\right) & -e^{-ik_y}\cos\left(k_x - \phi\right) \\ e^{ik_y}\cos\left(k_x - \phi\right) & -ie^{ik_y}\sin\left(k_x - \phi\right) \end{bmatrix} \begin{bmatrix} w_{k_x-\phi,k_y}^{(n)} + v_{k_x-\phi,k_y}^{(n)} \\ w_{k_x-\phi,k_y}^{(n)} - v_{k_x-\phi,k_y}^{(n)} \end{bmatrix}
$$

Focusing on the down channel, we know that at time step $n$, we have:

$$
\begin{aligned}
d_{k_x,k_y}^{(n)} &= 0.5\left(s_{k_x,k_y}^{(n)} - p_{k_x,k_y}^{(n)}\right) \\
&= 0.5\left(w_{k_x+n\phi,k_y}^{(n)} - v_{k_x+n\phi,k_y}^{(n)}\right)
\end{aligned}
$$

From this expression, we can obtain the following equation for $\langle y^2 \rangle_D$:

$$
\begin{aligned}
\langle y^2 \rangle_D &= \sum_{x,y} y^2 \left| D_{x,y}^{(n)} \right|^2 \\
&= \frac{1}{4\pi^2} \int_{k_y=0}^{2\pi} \int_{k_x=0}^{2\pi} d_{k_x,k_y}^{(n)*} \left(i\frac{d}{dk_y}\right)^2 d_{k_x,k_y}^{(n)} \, dk_x dk_y \\
&= \frac{1}{16\pi^2} \int_{k_y=0}^{2\pi} \int_{k_x=0}^{2\pi} \left(w_{k_x,k_y}^{(n)} - v_{k_x,k_y}^{(n)}\right)^* \left(i\frac{d}{dk_y}\right)^2 \left(w_{k_x,k_y}^{(n)} - v_{k_x,k_y}^{(n)}\right) \, dk_x dk_y
\end{aligned}
$$

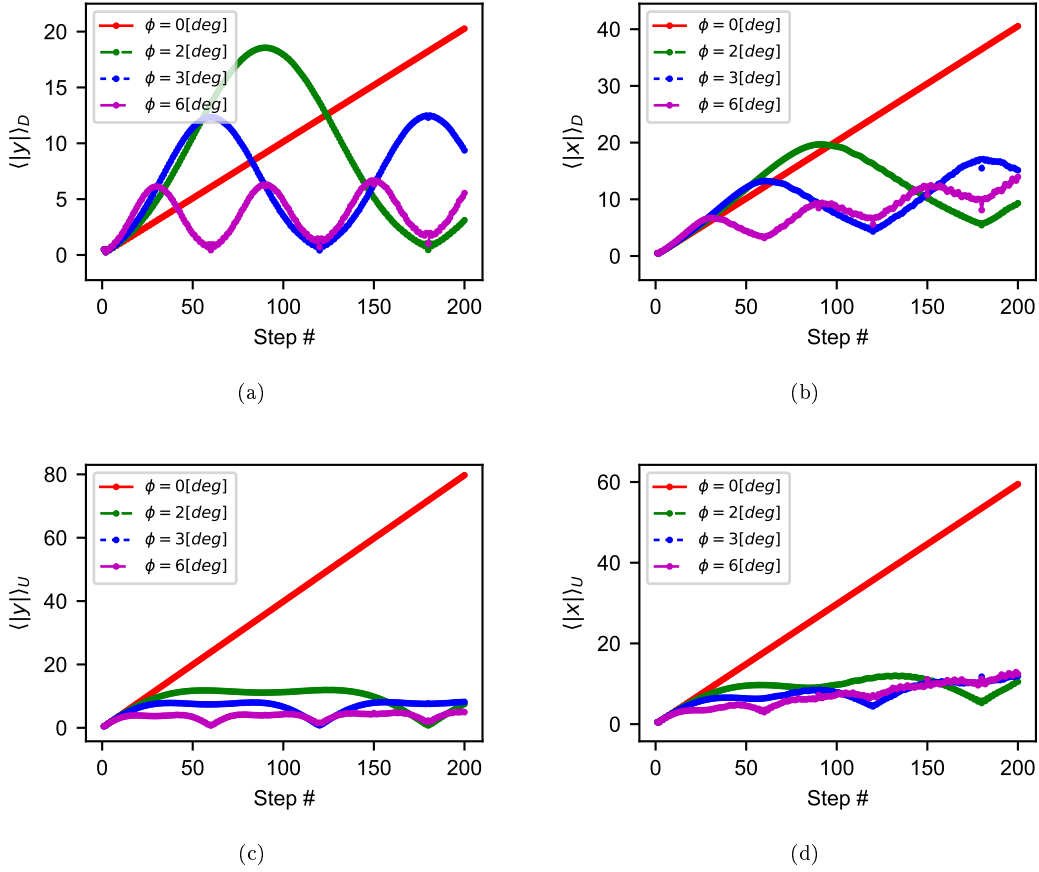

(a)

(b)

(c)

(d)

Figure S14. The norm one of (a) $y$ and (b) $x$ for the down channel as a function of the time step for different phase modulation values. The norm one of (c) $y$ and (d) $x$ for the up channel as a function of the time step for different phase modulation values.

Therefore, it is enough to obtain $w_{k_x,k_y}^{(n)} - v_{k_x,k_y}^{(n)}$ at time step $n$ in order to calculate the statistics of the quantum walk. Contrary to the zero-phase modulation, in which propagation matrices remain constant through time, here the evolution can be explained in terms of the product of $n$ matrices, which are not necessarily equal to each other:

$$
\begin{bmatrix} 0.5\left(w_{k_x,k_y}^{(n)} + v_{k_x,k_y}^{(n)}\right) \\ 0.5\left(w_{k_x,k_y}^{(n)} - v_{k_x,k_y}^{(n)}\right) \end{bmatrix} = \begin{bmatrix} ie^{-ik_y}\sin(k_x - \phi) & -e^{-ik_y}\cos(k_x - \phi) \\ e^{ik_y}\cos(k_x - \phi) & -ie^{ik_y}\sin(k_x - \phi) \end{bmatrix} \times \cdots
$$
$$
\times \begin{bmatrix} ie^{-ik_y}\sin(k_x - n\phi) & -e^{-ik_y}\cos(k_x - n\phi) \\ e^{ik_y}\cos(k_x - n\phi) & -ie^{ik_y}\sin(k_x - n\phi) \end{bmatrix} \begin{bmatrix} 1 \\ 0 \end{bmatrix}
$$

It is easy to verify that the expansion of $0.5\left(w_{k_x,k_y}^{(n)} - v_{k_x,k_y}^{(n)}\right)$ has terms proportional to $e^{ik_y n}$, $e^{ik_y(n-1)}$, ..., and $e^{ik_y(-n+2)}$. The first and last term represent the farthest $y$ position from the $x$ axis after $n$ steps. By obtaining the amplitudes of these two terms based on the above equation, we can expand $0.5\left(w_{k_x,k_y}^{(n)} - v_{k_x,k_y}^{(n)}\right)$ as:

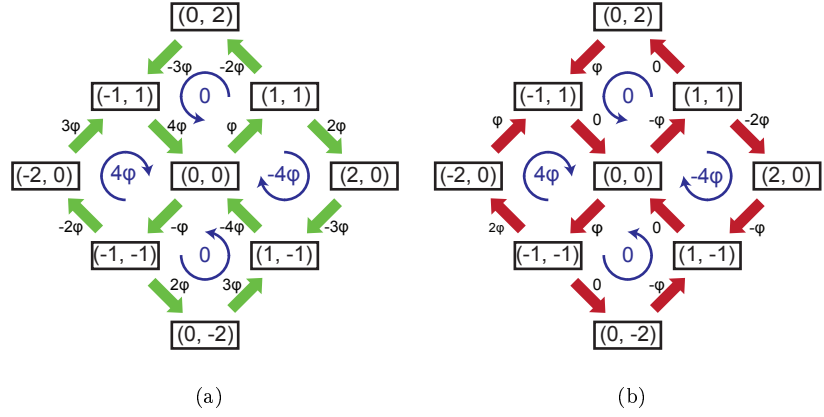

Figure S15. (a) The accumulated phases based on time-varying gauge fields implemented in this work along four sample closed paths. (b) The accumulated phases using a coordinate-dependent unitary operation along four similar closed paths.

$$0.5 \left( w^{(n)}_{k_x, k_y} - v^{(n)}_{k_x, k_y} \right) = (-i)^{n-1} e^{in k_y} \cos \left( k_x - n\phi \right) \prod_{p=1}^{n-1} \sin \left( k_x - p\phi \right)$$

$$+ \, i^{n-1} e^{-i(n-2)k_y} \cos \left( k_x - \phi \right) \prod_{p=2}^{n} \sin \left( k_x - p\phi \right)$$

$$+ \sum_{y=-n+3}^{n-1} a_y \left( k_x, \phi, n \right) e^{i k_y y}$$

Which means that

$$P_D \left( y = -n \right) = P_D \left( y = n - 2 \right)$$

$$= \frac{1}{2\pi} \int_{k_x = 0}^{2\pi} \cos^2 \left( k_x \right) \prod_{p=1}^{n-1} \sin^2 \left( k_x - p\phi \right) dk_x.$$

These expressions show the existence of zeros in the integrand at $k_x = p\phi$ for $p = 1$ to $n-1$. For $n \sim \frac{2\pi}{\phi}$, these zeros spread across the entire region of $k_x = 0$ to $k_x = 2\pi$. Especially, for the small values of $\phi$, the presence of these zeros causes that $P_D \left( y = -n \right)$ and $P_D \left( y = n - 2 \right)$ tend to zero at $n \sim \frac{2\pi}{\phi}$. For other values of $y \neq 0$, similar expressions can be obtained for $P_D \left( y \right)$, which they also become small for $\phi = \frac{2\pi}{n}$. This analysis predicts that $\langle y^2 \rangle_D$ tends toward zero for $n = 2\pi/\phi$.

Figures S13a and S13c summarize the numerical results for the variation of $\sqrt{\langle y^2 \rangle_D}$ and $\sqrt{\langle y^2 \rangle_U}$ with the time step for different values of $\phi$. The obtained results show that for $1 \ll n \ll \pi/\phi$, $\sqrt{\langle y^2 \rangle_D}$ varies as $n/\sqrt{6\pi}$, as one would expect from the zero phase modulation case. Moreover, the numerical results show that for small values of $\phi$, $\sqrt{\langle y^2 \rangle_D}$ reaches the maximum value at $n \cong \pi/\phi$. These results also show that $\langle y^2 \rangle_D$ tends toward zero at $n \cong 2\pi/\phi$, which is consistent with the above analysis. The effect of phase modulation on the quadratic mean of $x$ can be investigated as well. Numerical results as depicted in Figs. S13b and S13d show that the quadratic mean of $x$ also decreases relative to the zero-phase modulation case. However, the effect of phase modulation is more intense on decreasing the quadratic mean of $y$ as compared with the quadratic mean of $x$.

In addition to quadratic means, we can also look at the variations of the norm ones of the quantum walk distributions. Figures S14a and S14c summarize the numerical results for the variation of $\langle |y| \rangle_D$ and $\langle |y| \rangle_U$ with the time step for different values of $\phi$. The corresponding results for the $x$ direction are also shown in Figs. S14b and S14d.

### B. Creation of electric fields using various gauge fields:

In addition to the time-dependent approach, an electric field can be implemented through the use of a coordinate-dependent gauge field as well. In the latter case, instead of using a time-dependent gauge field, a phase modulation is applied that is not direction dependent and instead depends on the coordinate linearly. In this approach, an effective linear electric potential $V = -\mathcal{E}x$ is implemented that will lead to the generation of an electric field based on $\vec{\mathcal{E}} = -\nabla V$. The unitary operation in each time step has an extra term relative to the standard quantum walk evolution operator $U_0$ as $U_\phi = e^{i\phi x} U_0$. We can compare the effect of such a gauge field with the time-dependent gauge field considered in this work. In Fig. S15, we have considered four sample closed paths that start from the origin and return to it, and have calculated the total phase accumulated in them. As this figure shows, the net amount of phase in both approaches are similar, showing that indeed the time-varying gauge field will induce a similar phase accumulation in these closed paths as a conventional electric field in the $x$ direction. This similarity holds for any closed path starting from the origin and ending at it.

———————

[S1] H. Chalabi, S. Barik, S. Mittal, T. E. Murphy, M. Hafezi, and E. Waks, Physical Review Letters **123**, 150503 (2019).



\end{document}